\documentclass[reprint,nofootinbib,superscriptaddress]{revtex4-1} 
\newcommand{\SingleColFigScale}{0.95} 
\newcommand{\DoubleColFigScale}{0.95} 

\usepackage{graphicx}
\usepackage[usenames,dvipsnames]{xcolor}

\usepackage{amsmath}
\usepackage{amssymb}
\usepackage{mathtools}
\usepackage{accents}
\usepackage{bm}
\usepackage{dcolumn}
\usepackage{multirow}
\usepackage[pdfusetitle]{hyperref}
\definecolor{link}{rgb}{0.07, 0.07, 0.80}
\hypersetup{
    colorlinks=true,
    linkcolor=link,
    citecolor=link,
    urlcolor=link
}

\usepackage{txfonts}
\usepackage{wrapfig}

\makeatletter
\newcommand*{\nolink}[1]{{\begin{NoHyper}#1\end{NoHyper}}}
\newcommand{\manuallabel}[2]{\protect\def\@currentlabel{#2}\label{#1}\phantomsection}
\makeatother

\makeatletter
\renewcommand\frontmatter@abstractwidth{\dimexpr\textwidth-0.2in\relax}
\makeatother

\newenvironment{suppinfo}{
  \hspace{1ex}
  \section*{Supporting Information Available}
}{}

\begin{document}

\title{Low-Temperature Quantum Fokker-Planck and Smoluchowski Equations and\\ Their Extension to Multistate Systems}

\author{Tatsushi Ikeda}
\email{ikeda.tatsushi.37u@kyoto-u.jp}
\affiliation{Department of Chemistry, Graduate School of Science, Kyoto University, Kyoto 606-8502, Japan}

\author{Yoshitaka Tanimura}
\email{tanimura.yoshitaka.5w@kyoto-u.jp}
\affiliation{Department of Chemistry, Graduate School of Science, Kyoto University, Kyoto 606-8502, Japan}

\date{\today}




\begin{abstract}
  \begin{wrapfigure}{r}{\dimexpr3.5in-0.2in\relax}
    \flushleft
    \hspace{\dimexpr-0.4in\relax}
    \includegraphics[width=3.5in,height=1.375in]{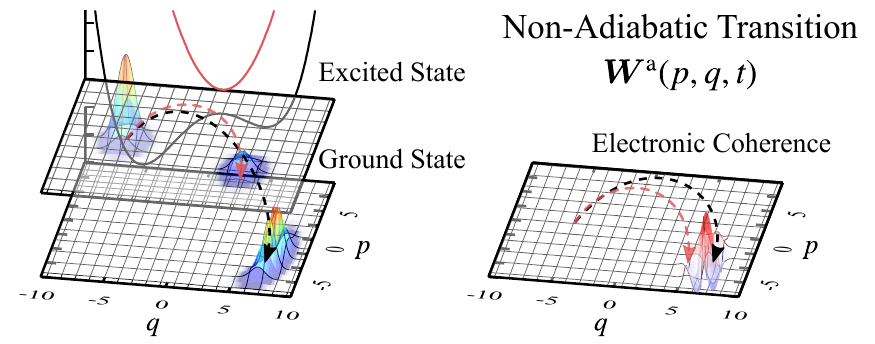}
  \end{wrapfigure}
  Simulating electron-nucleus coupled dynamics poses a non-trivial challenge and an important problem in the investigation of ultrafast processes involving coupled electronic and vibrational dynamics.
Because irreversibility of the system dynamics results from thermal activation and dissipation caused by the environment, in dynamical studies, it is necessary to include heat bath degrees of freedom in the total system.
  When the system dynamics involves high-energy electronic transitions, the environment is regarded to be in a low-temperature regime and we must treat it quantum mechanically.
  In this paper, we present rigorous and versatile approaches for investigating the dynamics of open systems with coupled electronic and vibrational degrees of freedom within a fully quantum mechanical framework.
  These approaches are based on a quantum Fokker-Planck equation and a quantum Smoluchowski equation employing a heat bath with an Ohmic spectral density, with non-Markovian low-temperature correction terms, and extensions of these equations to the case of multi-state systems.
  The accuracy of these equations was numerically examined for a single-state Brownian system, while their applicability was examined for multi-state double-well systems by comparing their results with those of the fewest-switch surface hopping and Ehrenfest methods with a classical Markovian Langevin force.
  Comparison of the transient absorption spectra obtained using these methods clearly reveals the importance of the quantum low-temperature correction terms.
  These equations allow us to treat non-adiabatic dynamics in an efficient way, while maintaining numerical accuracy.
  The C++ source codes that we developed, which allow for the treatment of the phase and coordinate space dynamics with any single-state or multi-state potential forms, are provided as Supporting Information.
\end{abstract}

\maketitle

\manuallabel{sec:s-derivation}{S1}
\manuallabel{sec:s-lt-qfpe}{S1.B}
\manuallabel{sec:s-lt-qse}{S1.C}
\manuallabel{sec:s-continuous-coordinate-representation}{S2}
\manuallabel{sec:s-langevin}{S3}
\manuallabel{sec:s-truncation}{S4}
\manuallabel{sec:s-uv-divergence}{S5}
\manuallabel{sec:s-numerical-details}{S6}
\manuallabel{sec:s-moyal-product}{S6.1}
\manuallabel{sec:s-trajectory-based-methods}{S6.2}

\section{INTRODUCTION}
Understanding non-adiabatic dynamics in electronic and bio-nanomaterials is fundamentally important in the study of many types of phenomena, ranging from photo-isomerization to spintronics.
Recent advances in experimental technologies have made it possible to observe such non-adiabatic processes that take place on very short timescales \cite{polli2010nature, iwamura2011jacs, lewis2016jpcl, dean2016chem, miyata2017mc}.
Theoretical input regarding the complex profiles of potential energy surfaces (PESs) and the non-adiabatic coupling among PESs is important for analyzing such ultrafast transport processes, in particular, those in materials involving biomolecular aggregates and crystalline solar cells \cite{prokhorenko2016jpcl, kato2018prl}.
For such systems, the surrounding molecules act as a heat bath and also play an essential role in determining the nature of the transport processes, because they either promote or suppress the wavepacket motion of the system through thermal activation and relaxation.

In the study of systems of the type considered here, while nuclear motion is often treated using a semiclassical approach, which is applicable to the case of heavy nuclei, non-adiabatic transition processes must be described using a purely quantum mechanical approach, because transitions between discretized electronic states are, in their essence, quantum dynamics.
For this reason, the effect of the environment should be treated using an open quantum model, even if the dynamics of the nuclei are semiclassical, because otherwise the quantum nature of the electronic transitions is not properly accounted for.
Indeed, ignoring the quantum effects of the environment, in particular in the low temperature regime, in which quantum effects become very important, leads to unphysical behavior.
For example, in the case that we employ a classical description of the environment in the low temperature regime, while electronic transitions and the motion of wavepackets are described by quantum mechanics, the positivity of the probability distributions of the electronic states cannot be maintained.
This is a fundamental complication, known as the ``positivity problem,'' which imposes a well-known limitation on the applicability of the quantum master equation without the rotating wave approximation \cite{dumcke1979zpb, pechukas1994prl}.
The positivity problem arises because the classical treatment of the environment leads to the violation of the quantum fluctuation-dissipation (QFD) theorem \cite{ford1988pra, frantsuzov1997cpl, dammak2009prl, tanimura2006jpsj}.

The excited state dynamics of systems exhibiting ultrafast coupled electronic and vibrational dynamical processes have been investigated with models that explicitly take into account nuclear degrees of freedom and electronic states through approaches employing equations of motion for wave functions, density matrices, phase space distributions \cite{thoss2000jcp, kuhl2002jcp, tanimura1994jcp, tanimura1997jcp, maruyama1998cpl, ikeda2017jcp, ikeda2018cp, kapral1999jcp, xie2014jcp}, and Gaussian quantum wavepackets \cite{ben1998jcp, ben2000jcpa, makhov2014jcp}, and approaches utilizing mixed quantum-classical trajectories \cite{tully1990jcp, coker1995jcp, subotnik2016arpc}.
However, many of these approaches were developed for isolated systems and were verified within the system which have a few degrees of freedom.
Moreover, varieties of assumptions (in particular, assumptions regarding the quantum dynamical treatment of the couplings between the electronic states and the nuclear coordinates) were introduced in such approaches and the assumptions severely limit their range of applicability.
Contrastingly, the multi-state quantum hierarchical Fokker-Planck equation (MS-QHFPE) approach, which is an extension of the quantum Fokker-Planck equation for the Wigner distribution function \cite{caldeira1983pa, tanimura1991pra, tanimura1992jcp, tanimura2015jcp} to multi-state systems \cite{tanimura1994jcp, tanimura1997jcp, maruyama1998cpl, ikeda2017jcp, ikeda2018cp} and is a variant of the hierarchical equations of motion (HEOM) theories \cite{tanimura1989jpsj, ishizaki2005jpsj, tanimura2006jpsj}, can treat any types of diabatic coupling and PES profiles with non-Markovian system-bath interactions described by a Drude spectral density.
However, although the MS-QHFPE approach allows us to compute the dynamics described by a multi-state system-bath Hamiltonian numerically rigorously, integrating the equations of motion is very computationally intensive, in particular, for a system described by multi-dimensional PESs.
Hence, presently, calculations carried out for two-dimensional systems are limited to the high-temperature Markovian case described by the MS-QFPE \cite{ikeda2018cp}.

While it has been found that non-Markovian effects arising from non-Ohmic environments are important in the description of exciton/electron transfer phenomena \cite{ishizaki2009jcp, fujihashi2015jcp, prokhorenko2016jpcl, kato2018prl}, the Ohmic heat bath model for nuclear dynamics has been (implicitly) employed in many investigations for models described by PESs that further coupled to a heat-bath, including a model that causes a Brownian/Drude spectral density after reducing the nuclear degrees of freedom \cite{leggett1984prb, garg1985jcp, tanimura1994jpsj, tanimura2012jcp}.
This results from the fact that the non-Markovian effects on the nuclear motion so far studied are regarded to be insignificant in such systems, in particular, when the damping on the nuclear motion is quite strong.
For this reason, although there have been several investigations employing Drude environments for the nuclear dynamics carried out on systems including nonlinear vibrational responses \cite{ishizaki2006jcp, sakurai2011jpca, ikeda2015jcp}, a ratchet system \cite{kato2013jpcb}, and a resonant tunneling diode system \cite{sakurai2013jpsj, sakurai2014njp}, in this paper, we derive equations of motion for single-state and multi-state systems employing the Ohmic environment:
low-temperature quantum Fokker-Planck equations (LT-QFPE) and low-temperature quantum Smoluchowski equations (LT-QSE) and their extensions to multi-state (MS) systems, MS-LT-QFPE and MS-LT-QSE.
As seen in the theory of quantum Brownian motion, within a quantum mechanical description, an Ohmic bath exhibits peculiar behavior in momentum space \cite{grabert1988pr, philip2000adp, ankerhold2001prl}.
We show that this difficulty can be avoided by properly treating the low temperature correction terms in the LT-QFPE and MS-LT-QFPE.
In the case of a heat bath with an Ohmic spectral density, the LT-QFPE and LT-QSE are sufficiently accurate, while also being sufficiently simple in comparison to the QHFPE.
These features make the LT-QFPE and LT-QSE suited for describing slowly decaying systems and systems rendered in multidimensional phase spaces.
Also, it is noteworthy that many of the existing formalisms, including those of the quantum Fokker-Planck equation \cite{caldeira1983pa, tanimura1991pra, tanimura1992jcp, tanimura2015jcp} and Zusman equation \cite{zusman1980cp, shi2009jcp2, shi2009jcp}, can be derived from the (MS-)LT-QFPE and (MS-)LT-QSE under certain conditions.

The organization of this paper is as follows.
In Sec.~\ref{sec:model}, we introduce a model with multiple electronic states described by the PESs that are coupled to a harmonic heat bath with an Ohmic spectral density.
Then, we present the MS-LT-QFPE and MS-LT-QSE and their single PES forms, LT-QFPE and LT-QSE.
In Sec.~\ref{sec:results}, we present numerical results for single-state Brownian and multi-state double-well systems to illustrate the validity and applicability of these approaches.
Section~\ref{sec:conclusion} is devoted to concluding remarks.
The C++ source codes that we developed are provided as Supporting Information.

\section{HAMILTONIAN AND FORMALISM}
\label{sec:model}
\subsection{Model}
Because the LT-QFPE and LT-QSE are the simpler, single-potential forms of the MS-LT-QFPE and MS-LT-QSE, we start with a multi-potential system.
We consider a molecular system with multiple electronic states $\{|j\rangle \}$ coupled to the nuclear coordinates.
For simplicity, we represent the nuclear coordinates by a single dimensionless coordinate, $q$.
Here and hereafter, we employ a dimensionless coordinate and a dimensionless momentum defined in terms of the actual coordinate and momentum, $\bar{q}$ and $\bar{p}$, as $q\equiv \bar{q}\sqrt {m\omega _{0}/\hbar }$ and $p\equiv \bar{p}/\sqrt {m\hbar \omega _{0}}$, where $\omega _{0}$ is the characteristic vibrational frequency of the system and $m$ is the effective mass.
The reaction coordinate is also bilinearly coupled to the harmonic bath coordinates, $\vec{x}\equiv (\dots ,x_{\xi },\dots )$.
The Hamiltonian of the total system is expressed as \cite{caldeira1983pa}
\begin{align}
  \hat{H}_{\mathrm{tot}}(p,q;\vec{p},\vec{x})&\equiv \hat{H}(p,q)+\hat{H}_{\mathrm{B}}(\vec{p},\vec{x};q),
  \label{eq:total-Hamiltonian}
\end{align}
where the system Hamiltonian, $\hat{H}(p,q)$, is defined as
\begin{align}
  \hat{H}(p,q)\equiv \frac{\hbar \omega _{0}}{2}\hat{p}^{2}+\sum _{j,k}|j\rangle U_{jk}^{\mathrm{d}}(\hat{q})\langle k|.
  \label{eq:system-Hamiltonian}
\end{align}
Here, the nuclear and electronic operators are denoted by hats, and the direct products with the unit operator in the kinetic and bath terms ($\otimes \hat{1}$) are omitted.
The diagonal element $U_{jj}^{\mathrm{d}}(q)$ is the diabatic PES of $|j\rangle $, and the off-diagonal element $U_{jk}^{\mathrm{d}}(q)$ $(j{\neq }k)$ represents the diabatic coupling between $|j\rangle $ and $|k\rangle $.
The vibrational frequency $\bar{\omega }$ at a local minimum of the potential $q_{0}$ is determined by the curvature of the PESs as
\begin{align}
  \hbar \bar{\omega }\simeq \sqrt {\hbar \mathstrut \omega _{0}\left.\frac{\partial ^{2}}{\partial q^{2}}U_{j_{0}j_{0}}(q)\right|_{q=q_{0}}},
\end{align}
where $j_{0}$ is a primary state of the vibrational dynamics.
Therefore, the frequency $\mathstrut \omega _{0}$ is chosen to be $\hbar \omega _{0}\simeq \partial ^{2}U_{j_{0}j_{0}}(q)/\partial q^{2}|_{q=q_{0}}$ in order to have $\bar{\omega }\simeq \omega _{0}$.
The bath Hamiltonian $\hat{H}_{\mathrm{B}}(\vec{p},\vec{x};q)$ is defined as
\begin{align}
  \hat{H}_{\mathrm{B}}(\vec{p},\vec{x};q)&\equiv \sum _{\xi }\frac{\hbar \omega _{\xi }}{2}\left[\hat{p}_{\xi }^{2}+\left(\hat{x}_{\xi }-\frac{{g_{\xi }}}{{\omega _{\xi }}}\hat{q}\right)^{2}\right],
  \label{eq:environment-Hamiltonian}
\end{align}
where $\omega _{\xi }$, $p_{\xi }$, and $g_{\xi }$ are the vibrational frequency, conjugate momentum, and system-bath coupling constant of the $\xi $th dimensionless bath mode, $x_{\xi }$.
\begin{subequations}
  The bath is characterized by the dissipation and fluctuation that it engenders.
  These are represented by the relaxation function
  \begin{align}
    R(t)=\frac{2}{\pi }\int _{0}^{\infty }\!d\omega \,\frac{\mathcal{J}(\omega )}{\omega }\cos \omega t
    \label{eq:rlx-exact}
  \end{align}
  and the symmetrized correlation function
  \begin{align}
    C(t)=\frac{2}{\pi }\int _{0}^{\infty }d\omega \mathcal{J}(\omega )\left(n(\omega )+\frac{1}{2}\right)\cos \omega t
    \label{eq:sym-exact},
  \end{align}
  where the spectral density is defined as $\mathcal{J}(\omega )\equiv \pi \sum _{\xi }(g_{\xi }^{2}/2)\delta (\omega -\omega _{\xi })$, and we have introduced the Bose-Einstein distribution function, $n(\omega )\equiv (e^{\beta \hbar \omega }-1)^{-1}$, for the inverse temperature divided by the Boltzmann constant, $\beta \equiv 1/k_{\mathrm{B}}T$.
\end{subequations}

We choose the coefficients $\nu _{k}$ and $\eta _{k}$ so as to realize the relation
\begin{align}
  n(\omega )+\frac{1}{2}\simeq \frac{1}{\beta \hbar }\frac{1}{\omega }+\sum ^{K}_{k}\frac{2\eta _{k}}{\beta \hbar }\frac{\omega }{\omega ^{2}+\nu _{k}^{2}}
  \label{eq:temperature-kernel}
\end{align}
for finite $K$, where the first term on the right-hand side is the classical contribution from the temperature, and the remaining terms are the quantum low-temperature (QLT) corrections.
The Matsubara decomposition scheme (MSD) can be applied straightforwardly to the above.
In this scheme, we set $\eta _{k}=1$ and $\nu _{k}=\tilde{\nu }_{k}$, where $\tilde{\nu }_{k}\equiv 2\pi k/\beta \hbar $ is the $k$th Matsubara frequency \cite{tanimura1990pra, ishizaki2005jpsj}.
In this paper, we employ the Pad\'e spectral decomposition $[N{-}1/N]$ (PSD$[N{-}1/N]$) scheme to enhance the computational efficiency while maintaining the accuracy \cite{hu2010jcp, hu2011jcp, ding2012jcp}.

In order to reduce the computational times for the computations of the non-adiabatic dynamics with any forms of the PESs, here we employ an Ohmic spectral density, expressed as
\begin{align}
  \mathcal{J}(\omega )&=\frac{\zeta }{\omega _{0}}\omega ,
  \label{eq:ohmic-density}
\end{align}
where $\zeta $ is the system-bath coupling strength.
Then we have
\begin{subequations}
  \begin{align}
    &R(t)=\frac{\zeta }{\omega _{0}}\cdot 2\delta (t)
    \label{eq:rlx-approx}
    \intertext{and}
    &\begin{aligned}
       C(t)\simeq C_{K}(t)&=\frac{\zeta }{\omega _{0}}\left(\frac{1}{\beta \hbar }+\sum _{k}^{K}\frac{2\eta _{k}}{\beta \hbar }\right)\cdot 2\delta (t)\\
       &\quad -\sum _{k}^{K}\frac{\zeta }{\omega _{0}}\frac{2\eta _{k}}{\beta \hbar }\cdot \nu _{k}e^{-\nu _{k}\left|t\right|}.
     \end{aligned}
    \label{eq:sym-approx}
  \end{align}
\end{subequations}  
In the case of a harmonic PES with frequency $\omega _{0}$, the conditions $\zeta <2\omega _{0}$, $\zeta =2\omega _{0}$, and $\zeta >2\omega _{0}$ correspond to the underdamped, critically damped, and overdamped cases, respectively.
\begin{figure}
  \centering
  \includegraphics[scale=\SingleColFigScale]{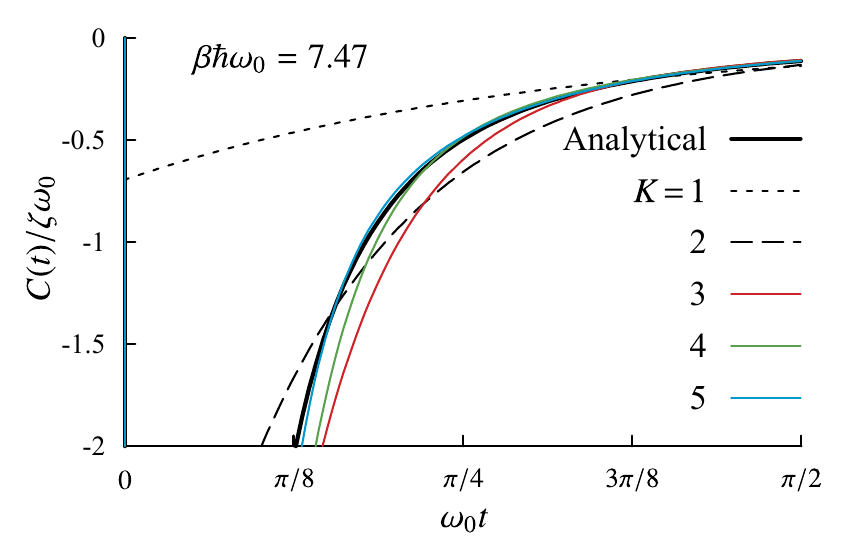}
  \caption{ The symmetrized correlation function, $C(t)$, for an Ohmic spectral density, Eq.~\eqref{eq:ohmic-density}, at temperature $\beta \hbar \omega _{0}=7.47$, which is in the low-temperature regime.
    The thick black solid curve represents the exact expression, Eq.~\eqref{eq:sym-exact}, and the other curves represent $C_{K}(t)$, Eq.~\eqref{eq:sym-approx}, for various values of the cutoff, $K$.
    The values of the coefficients $\nu _{k}$ and $\eta _{k}$ for each value of $K$ are given in Table~\ref{tab:psd_coeffs}.
    The fast decay components in the exact expression and the delta-function components in Eq.~\eqref{eq:sym-approx}, which are overlapped with the $t=0$ axis, are not displayed.
  }
  \label{fig:bath_correlation}
\end{figure}
\begin{table}
  \centering
  \caption{
    PSD[N-1/N] coefficients used for $K=1$--$5$.
  }
  \begin{tabular}{c|rrr}
    \hline\hline
    $K$ & $k$ & \multicolumn{1}{c}{$\beta \hbar \nu _{k}$} & \multicolumn{1}{c}{$\eta _{k}$} \\
    \hline
    $1$ & $1$ & $7.745967$ &  $2.5\phantom{00000}$ \\
    \hline
    \multirow{2}{*}{2} & $1$ & $6.305939$ & $1.032824$ \\
    & $2$ & $19.499618$ &  $5.967176$ \\
    \hline
    \multirow{3}{*}{3} & $1$ & $6.2832903$ & $1.000227$ \\
    & $2$ & $12.9582867$ & $1.300914$ \\
    & $3$ & $36.1192894$ & $11.198859$ \\
    \hline
    \multirow{4}{*}{4} & $1$ & $6.283185$  & $1.000000$ \\
    & $2$ & $12.579950$   & $1.015314$ \\
    & $3$ & $20.562598$   & $1.905605$ \\
    & $4$ & $57.787940$    & $18.079081$ \\
    \hline
    \multirow{5}{*}{5} & $1$ & $6.283185$ & $1.000000$ \\
    & $2$ & $12.566542$   & $1.000262$ \\
    & $3$ & $19.004690$ & $1.113033$ \\
    & $4$ & $29.579276$ & $2.800147$ \\
    & $5$ & $84.536926$ & $26.586558$ \\
    \hline\hline
  \end{tabular}
  \label{tab:psd_coeffs}
\end{table}
In Fig.~\ref{fig:bath_correlation}, we plot $C_{K}(t)$ for various values of the cutoff $K$, at the temperature $\beta \hbar \omega _{0}=7.47$.
As this figure and Eq.~\eqref{eq:sym-approx} indicate, the fluctuation term is always non-Markovian due to the quantum nature of the noise, and it can be regarded as Markovian only in the high-temperature limit, $\beta \hbar \omega _{\mathrm{0}}\ll 1$, in which the heat bath exhibits classical behavior \cite{tanimura2006jpsj, weiss2011book}.
This is an important conclusion obtained from the QFD theorem, namely, that the negative non-Markovian terms appear in the case that we do not use a time-coarse-grained, Markovian description.
Note that, in the case that we employ an Ohmic spectral density without cutoff functions (e.g.~Lorentzian cutoff and exponential cutoff), some of the physical observables, including the mean square of the momentum, $\langle p^{2}\rangle $, diverge due to the divergence of the first and second terms in Eq.~\eqref{eq:sym-approx} under the infinite summation of the Matsubara frequencies, if there is no finite cutoff function.
This divergence is often referred to as the ultra-violet divergence \cite{grabert1988pr, philip2000adp, ankerhold2001prl}.
In practice, we can ignore QLT correction terms whose frequencies are sufficiently greater than the characteristic frequency of the system, because the random force generated by such terms is averaged out over a sufficiently short timescale that its influence on the dynamics of interest is negligible.
In this way, we are able to calculate non-diverging physical observables, e.g.~ the mean square of the coordinate, $\langle q^{2}\rangle $, by simply ignoring the contribution from the high-frequency QLT correction terms by implementing the cutoff $K$, while diverging physical observables still tend to diverge.

\subsection{Multi-State Low-Temperature Quantum Fokker-Planck Equations}
\label{sec:heom}
The state of the total system is represented by the density operator, $\hat{\rho }_{\mathrm{tot}}(z,z',\vec{x},\vec{x}\,')\equiv \langle \vec{x}|\langle z|\hat{\rho }_{\mathrm{tot}}(t)|z'\rangle |\vec{x}\,'\rangle $, where $|z\rangle $ and $|\vec{x}\rangle $ are the eigenstate of the system and bath coordinate operators, respectively.
We consider the reduced density matrix in the diabatic representation of the system subspace, defined as
\begin{align}
  \rho _{jk}^{\mathrm{d}}(z,z',t)&\equiv \langle j|\hat{\rho }(z,z',t)|k\rangle ,
\end{align}
where $\hat{\rho }(z,z',t)\equiv \mathrm{Tr}_{\mathrm{B}}\{\hat{\rho }_{\mathrm{tot}}(z,z',\vec{x},\vec{x}',t)\}$ is the reduced density operator and $\mathrm{Tr}_{\mathrm{B}}\{\dots \}\equiv \int \!d\vec{x}\,\int \!d\vec{x}\,'\delta (\vec{x}-\vec{x}\,')\{\dots \}$ represents the trace operation in the bath subspace.
The diagonal and off-diagonal elements, $\rho ^{\mathrm{d}}_{jj}$ and $\rho ^{\mathrm{d}}_{jk}$ ($j{\neq }k$), represent the population of $|j\rangle $ and the coherence between $|j\rangle $ and $|k\rangle $, respectively.
Hereafter, we employ the matrix forms of the reduced density matrix and the diabatic PESs: $\{\bm{\rho }^{\mathrm{d}}(z,z')\}_{jk}\equiv \rho ^{\mathrm{d}}_{jk}(z,z')$ and $\{\bm{U}^{\mathrm{d}}(z)\}_{jk}\equiv U^{\mathrm{d}}_{jk}(z)$.

We now introduce the Wigner distribution function, which is the quantum analogy of the classical distribution function in phase space.
For a multi-state system, the multi-state Wigner distribution function (MS-WDF) is defined as \cite{tanimura1994jcp, tanimura1997jcp, maruyama1998cpl, ikeda2017jcp, ikeda2018cp}
\begin{align}
  \bm{W}^{\mathrm{d}}(p,q,t)&\equiv \frac{1}{2\pi }\int \!dr\,e^{-ipr}\bm{\rho }^{\mathrm{d}}\left(q+\frac{r}{2},q-\frac{r}{2}\right),
\end{align}
where $q\equiv (z+z')/2$ and $r\equiv z-z'$.
Both $p$ and $q$ are now c-numbers in this phase space representation.

The reduced dynamics of $\bm{\rho }^{\mathrm{d}}(z,z',t)$ and $\bm{W}^{\mathrm{d}}(p,q,t)$ are expressed in the path integral framework using the Feynman-Vernon influence functional \cite{feynman1963ap}.
Their time evolutions can be described by a set of time differential equations in the HEOM form (see Sec.~\nolink{\ref{sec:s-derivation}} in the Supporting Information).
In the present case, these equations are the following:
\begin{widetext}
  \begin{align}
    \begin{split}
      \frac{\partial }{\partial t}\bm{W}_{\vec{n}}^{\mathrm{d}}(p,q,t)
      &=-\left(\mathcal{L}_{\mathrm{qm}}^{\mathrm{d}}(p,q)+\sum _{k}^{K}n_{k}\nu _{k}+\hat{\Xi }_{K}^{\mathrm{d}}(p,q)\right)\bm{W}_{\vec{n}}^{\mathrm{d}}(p,q,t)\\
      &\quad -\sum _{k}^{K}\hat{\Phi }^{\mathrm{d}}(p,q)\bm{W}_{\vec{n}+\vec{e}_{k}}^{\mathrm{d}}(p,q,t)
      -\sum _{k}^{K}n_{k}\nu _{k}\hat{\Theta }_{k}^{\mathrm{d}}(p,q)\bm{W}_{\vec{n}-\vec{e}_{k}}^{\mathrm{d}}(p,q,t),
    \end{split}
    \label{eq:lt-qfpe-d}
  \end{align}
\end{widetext}
where $\vec{n}\equiv (\dots ,n_{k},\dots )$ is a $K$-dimensional multi-index whose components are all non-negative integers and $\vec{e}_{k}\equiv (0,\dots ,1,0,\dots )$ is the $k$th unit vector.
The multi-index $\vec{n}$ represents the index of the hierarchy, and physically, the first hierarchical element, $\bm{W}_{\vec{0}}^{\mathrm{d}}(p,q,t)$, corresponds to the MS-WDF, $\bm{W}^{\mathrm{d}}(p,q,t)$.
The rest of the hierarchical elements serve only to facilitate treatment of the non-Markovian system-bath interaction that arises from the QLT effects.

\begin{subequations}
  The quantum Liouvillian for the MS-WDF is given by
  \begin{align}
    \mathcal{L}_{\mathrm{qm}}^{\mathrm{d}}(p,q)&\equiv \mathcal{K}(p,q)+\mathcal{U}_{\mathrm{qm}}^{\mathrm{d}}(p,q),
    \label{eq:qm-Liouvillian}
  \end{align}
  where
  \begin{align}
    &\begin{aligned}
      \mathcal{K}(p,q)\bm{W}(p,q)&\equiv \omega _{0}p\frac{\partial }{\partial q}\bm{W}(p,q)\\
    \end{aligned}
    \intertext{and}
    &\begin{aligned}
       \mathcal{U}_{\mathrm{qm}}^{\mathrm{d}}(p,q)\bm{W}(p,q)
       &\equiv \frac{i}{\hbar }\Bigl(\bm{U}^{\mathrm{d}}(q){\star }\bm{W}(p,q)-\bm{W}(p,q){\star }\bm{U}^{\mathrm{d}}(q)\Bigr)
       \label{eq:qm-potential-term}
    \end{aligned}
  \end{align}
  are the kinetic and potential terms in the diabatic representation, respectively.
\end{subequations}
Here, we have introduced the star operator, $\star $, which represents the Moyal product, defined as \cite{moyal1949mpcps, imre1967jmp}
\begin{align}
  \star &\equiv \exp \left[{\frac{i}{2}\Bigl(\underaccent{\leftarrow }{\partial }_{q}\underaccent{\rightarrow }{\partial }_{p}-\underaccent{\rightarrow }{\partial }_{q}\underaccent{\leftarrow }{\partial }_{p}\Bigr)}\right].
  \label{eq:moyal-star}
\end{align}
The differentiation operators from the left and right appearing here are defined as
\begin{align}
  \underaccent{\rightarrow }{\partial }_{x}f(x)=f(x)\underaccent{\leftarrow }{\partial }_{x}&\equiv \frac{\partial f(x)}{\partial x}.
\end{align}
\begin{subequations}
  The operators for the fluctuation and dissipation, $\hat{\Phi }^{\mathrm{d}}(p,q)$, $\hat{\Theta }_{k}^{\mathrm{d}}(p,q)$, and $\hat{\Xi }_{k}^{\mathrm{d}}(p,q)$, appearing in Eq.~\eqref{eq:lt-qfpe-d}, are defined as
  \begin{align}
    \hat{\Phi }^{\mathrm{d}}(p,q)&\equiv -\frac{\partial }{\partial p},\\
    \hat{\Theta }_{k}^{\mathrm{d}}(p,q)&\equiv \frac{\zeta }{\omega _{0}}\frac{2\eta _{k}}{\beta \hbar }\frac{\partial }{\partial p},
    \intertext{and}
    \begin{split}
      \hat{\Xi }_{K}^{\mathrm{d}}(p,q)&\equiv -\frac{\zeta }{\omega _{0}}\frac{\partial }{\partial p}\biggl(\omega _{0}p+\frac{1}{\beta \hbar }\frac{\partial }{\partial p}\biggr)
      +\sum _{k}^{K}\hat{\Phi }^{\mathrm{d}}(p,q)\hat{\Theta }_{k}^{\mathrm{d}}(p,q).
    \end{split}
    \label{eq:xi-d}
  \end{align}
  The first two operators in the above equations, $\hat{\Phi }^{\mathrm{d}}(p,q)$ and $\hat{\Theta }_{k}^{\mathrm{d}}(p,q)$, arises from Eq.~\eqref{eq:rlx-approx} and the first term in Eq.~\eqref{eq:sym-approx}, while the last operator, $\hat{\Xi }_{K}^{\mathrm{d}}(p,q)$, arises from the second term in Eq.~\eqref{eq:sym-approx}.
\end{subequations}
The derivation of Eq.~\eqref{eq:lt-qfpe-d} is presented in Sec.~\nolink{\ref{sec:s-lt-qfpe}} of the Supporting Information.
Because Eq. ~\eqref{eq:lt-qfpe-d} is a generalization of the multi-state quantum Fokker-Planck equation (MS-QFPE) \cite{tanimura1994jcp, tanimura1997jcp, maruyama1998cpl, ikeda2017jcp} valid in the low-temperature regime, we refer to these equations as the multi-state low-temperature quantum Fokker-Planck equations (MS-LT-QFPE).

For a single-state system, the matrices $\bm{W}^{\mathrm{d}}(p,q,t)$ and $\bm{U}^{\mathrm{d}}(q)$ reduce to scalar functions, $W(p,q,t)$ and $U(q)$.
In this case, we refer to Eq.~\eqref{eq:lt-qfpe-d} as the low-temperature quantum Fokker-Planck equations (LT-QFPE).
These equations can be understood as an extension of the quantum Fokker-Planck equation (QFPE) \cite{caldeira1983pa, tanimura1991pra, tanimura1992jcp, tanimura2015jcp}.

The conventional (multi-state) quantum hierarchical Fokker-Planck equations ((MS-)QHFPE) with a Drude spectral density, $\mathcal{J}^{\mathrm{D}}(\omega )\propto \omega \gamma _{\mathrm{D}}^{2}/(\omega ^{2}+\gamma _{\mathrm{D}}^{2})$, where $\gamma _{\mathrm{D}}$ is the cutoff frequency, are capable of treating systems subject to non-Markovian noise, and are not limited to the case of an Ohmic spectral density \cite{sakurai2011jpca, kato2013jpcb, sakurai2014njp, tanimura2015jcp}.
However, the (MS-)QHFPE require a ($K+1$)-dimensional multi-index $\vec{n}'\equiv (n_{0},\vec{n})$ (i.e.~the additional index $n_{0}$) to describe non-Markovian dynamics caused by a finite value of $\gamma _{\mathrm{D}}$, and therefore computationally more expensive than the (MS-)LT-QFPE.
Moreover, the (MS-)QHFPE become unstable in the Ohmic limit (i.e.~$\gamma _{\mathrm{D}}\gg 1$) at low temperatures due to the fast decaying terms with $\gamma _{\mathrm{D}}$, while the (MS-)LT-QFPE is sufficiently accurate and also being sufficiently simple in comparison to the (MS-)QHFPE.
Thus, although applicability of these equations is limited to the Ohmic case, the computational cost to solve the (MS-)LT-QFPE is suppressed than that to solve the (MS-)QHFPE.
These features make the (MS-)LT-QFPE and (MS-)LT-QSE suited for describing slowly decaying systems and systems rendered in multidimentional phase spaces.
Note that in the case that the diabatic PESs of the system are harmonic, the MS-LT-QFPE yields the same results as the HEOM for a reduced electronic system with a Brownian spectral density \cite{tanimura1994jpsj, tanaka2009jpsj}.
In Appendices \ref{sec:appendx-continuous} and \ref{sec:appendx-langevin}, we present a stochastic Liouville description of the (MS-)LT-QFPE and Langevin description of the LT-QFPE, respectively.

\subsection{Multi-State Low-Temperature Quantum Smoluchowski Equations}
In this section, we present the asymptotic form of Eq.~\eqref{eq:lt-qfpe-d} in the Smoluchowski limit, i.e.~in the case $\zeta \gg \omega _{0}~\text{and}~\omega _{e}$, where $\omega _{e}$ is the characteristic frequency of the electronic transition dynamics.
We introduce the following probability distribution in coordinate space:
\begin{align}
  \bm{f}^{\mathrm{d}}(q,t)&\equiv \int \!dp\,\bm{W}^{\mathrm{d}}(p,q,t).
\end{align}
In the Smoluchowski limit, the equations of motion for $\bm{f}^{\mathrm{d}}(q)$ are
\begin{widetext}
  \begin{align}
    \begin{split}
      \frac{\partial }{\partial t}\bm{f}_{\vec{n}}^{\mathrm{d}}(q,t)
      &=-\left[\mathcal{E}^{\mathrm{d}}(q)+\sum _{k}^{K}n_{k}\nu _{k}+\frac{\omega _{0}}{\zeta }\left(\mathcal{F}^{\mathrm{d}}(q)+\hat{\Xi }_{K}^{\mathrm{od,d}}(q)\right)\right]\bm{f}_{\vec{n}}^{\mathrm{d}}(q,t)\\
      &\quad -\sum _{k}^{K}\hat{\Phi }^{\mathrm{od,d}}(q)\bm{f}_{\vec{n}+\vec{e}_{k}}^{\mathrm{d}}(q,t)
      -\frac{\omega _{0}}{\zeta }\sum _{k}^{K}n_{k}\nu _{k}\hat{\Theta }_{k}^{\mathrm{od},\mathrm{d}}(q)\bm{f}_{\vec{n}-\vec{e}_{k}}^{\mathrm{d}}(q,t),
    \end{split}
    \label{eq:lt-qse-d}
  \end{align}
\end{widetext}
\begin{subequations}
  where
  \begin{align}
    \mathcal{E}^{\mathrm{d}}(q)\bm{f}(q,t)&\equiv \frac{i}{\hbar }\bigl(\bm{U}^{\mathrm{d}}(q)\bm{f}(q,t)-\bm{f}(q,t)\bm{U}^{\mathrm{d}}(q)\bigr)
  \end{align}
  corresponds to the Liouville-von Neumann equation for the electronic subspace and
  \begin{align}
    \mathcal{F}^{\mathrm{d}}(q)\bm{f}(q,t)&\equiv \frac{\partial }{\partial q}\frac{1}{2}\Bigl(\bm{F}^{\mathrm{d}}(q)\bm{f}(q,t)+\bm{f}(q,t)\bm{F}^{\mathrm{d}}(q)\Bigr)
  \end{align}
  is the drift term that arises from the force $\bm{F}^{\mathrm{d}}(q)\equiv -(1/\hbar )\partial \bm{U}^{\mathrm{d}}(q)/\partial q$.
\end{subequations}
\begin{subequations}
  The operators 
  \begin{align}
    \hat{\Phi }^{\mathrm{od},\mathrm{d}}(q)&=-\frac{\partial }{\partial q}\\
    \intertext{and}
    \hat{\Theta }_{k}^{\mathrm{od},\mathrm{d}}(q)&=\frac{2\eta _{k}}{\beta \hbar }\frac{\partial }{\partial q}
  \end{align}
  represent the non-Markovian parts of the noise, while
  \begin{align}
    \hat{\Xi }_{K}^{\mathrm{od},\mathrm{d}}(q)&=-\frac{1}{\beta \hbar }\frac{\partial ^{2}}{\partial q^{2}}+\sum _{k}^{K}\hat{\Phi }^{\mathrm{od},\mathrm{d}}(q)\hat{\Theta }_{k}^{\mathrm{od},\mathrm{d}}(q)
  \end{align}
  represent the Markovian part of the noise.
  The superscript ``od'' means ``overdamped''.
\end{subequations}
The derivation of Eq.~\eqref{eq:lt-qse-d} is given in Sec.~\nolink{\ref{sec:s-lt-qse}} of the Supporting Information.
In the case of a single-state system, the matrices $\bm{f}^{\mathrm{d}}(q,t)$, $\bm{U}^{\mathrm{d}}(q)$ and $\bm{F}^{\mathrm{d}}(q)$ reduce to scalar functions, $f(q,t)$, $U(q)$ and $F(q)$, respectively.
The relationship between Eq.~\eqref{eq:lt-qfpe-d} and Eq.~\eqref{eq:lt-qse-d} is similar to the relationship between the Fokker-Planck (Kramers) equation and the Smoluchowski equation \cite{davies1954pr, risken1989book}.
For this reason, we refer to Eq.~\eqref{eq:lt-qse-d} as the (multi-state) low-temperature quantum Smoluchowski equations ((MS-)LT-QSE), while we refer to this as the (multi-state) Smoluchowski equation ((MS-)SE) in the high-temperature limit.

A quantum mechanical extension of the Smoluchowski equation valid in the low-temperature regime is known as the quantum Smoluchowski equation (QSE), which treats QLT effects in the framework of the Markovian approximation \cite{ankerhold2007book, philip2000adp, ankerhold2001prl, ankerhold2008prl, maier2010pre, maier2010pre2}.
However, because QLT corrections are in principle non-Markovian as shown in Eq.~\eqref{eq:sym-approx}, when we lower the bath temperature or we study a system with high energy, the QSE becomes inaccurate.
Contrastingly, the (MS-)LT-QSE can describe non-Markovian terms that is necessary to satisfy the QFD theorem, the (MS-)LT-QSE is applicable to a wider range of physical conditions than the QSE, as shown in Appendix \ref{sec:appendx-qse}.
For electron transfer problems with harmonic PESs, an extension of the Smoluchowski equation to multi-state systems has been carried out as the Zusman equation (ZE) \cite{zusman1980cp, srabani1994jcp, zusman1996jcp, shi2009jcp2}.
However, the original ZE theory does not treat quantum dynamical effects from electronic states properly, as shown in Appendix \ref{sec:appendx-Zusman}.
The MS-SE can be regarded as a generalization of the ZE for arbitrary PESs with describing the quantum dynamical effects from the electric states accurately.
The MS-LT-QSE can be regarded as a generalization of the MS-SE with the QLT correction terms.
 Several extensions of the ZE theory valid in the low-temperature regime have been carried out as the following:
The generalized ZE is constructed as an extension of the quantum Smoluchowski equation to multi-state systems \cite{ankerhold2004jcp}.
In the modified ZE theory \cite{shi2009jcp} and stochastic ZE theory \cite{lili2013njp}, the effects of non-Markovian QLT terms are incorporated using an integral-differential equation similar to that used in second-order perturbation theories and using a stochastic differential equation, respectively.
Contrastingly, the MS-LT-QSE is a non-Markovian, non-perturbative, and deterministic approach in the framework of the HEOM formalism.
Note that when the diabatic PESs of the system are harmonic, the MS-LT-QSE gives the same result as the HEOM for a reduced electronic system with an overdamped Brownian/Drude spectral density \cite{tanimura1994jpsj, ishizaki2009jcp}.
In Appendices \ref{sec:appendx-continuous} and \ref{sec:appendx-langevin}, we also present a stochastic Liouville description of the (MS-)LT-QSE and Langevin description of the LT-QSE, respectively.

\section{NUMERICAL RESULTS}
\label{sec:results}
In principle, with the (MS-)LT-QFPE, we are able to calculate various physical quantities with any desired accuracy by adjusting the number of low temperature corrections terms, while the (MS-)LT-QSE is sufficient for computing physical quantities under overdamped conditions.
Here, we first examine the validity of Eqs.~\eqref{eq:lt-qfpe-d} and ~\eqref{eq:lt-qse-d} by presenting the results obtained from numerical integrations of these equations in the case of a Brownian oscillator, for which exact solutions are known.
Then, we demonstrate the applicability of these equations by using them to compute the population dynamics and transient absorption spectrum for a multi-state double-well system.

\subsection{Single-State Case: Brownian Oscillator}
\label{sec:brownian}
Here, we consider the case of a single PES described by the harmonic potential
\begin{align}
  U(q)&=\frac{\hbar \omega _{0}}{2}q^{2}.
  \label{eq:harmonic-PES}
\end{align}
The validity of the reduced equation of motion for a non-Markovian system can be examined by comparing the results obtained from a set of numerical tests (non-Markovian tests) to the analytically derived solution in the case of the Brownian oscillator \cite{tanimura2015jcp}.

First, we study the equilibrium distribution function in the case of the above harmonic potential.
For this PES, we have the following analytical expression for the equilibrium distribution \cite{grabert1988pr, weiss2011book}:
\begin{align}
  f(q)&=\frac{1}{\sqrt {\mathstrut 2\pi \langle q^{2}\rangle _{\beta ,\zeta }}}e^{-q^{2}/2\langle q^{2}\rangle _{\beta ,\zeta }}.
  \label{eq:analytical-distribution}
\end{align}
Here,
\begin{subequations}
  \begin{align}
    \langle q^{2}\rangle _{\beta ,\zeta }&=\frac{2}{\pi }\int _{0}^{\infty }\!d\omega \,\tilde{C}_{q}(\omega )\\
    \label{eq:analytical-q-variance}
    &=\frac{\omega _{0}}{\beta \hbar }\sum _{k=-\infty }^{\infty }\frac{1}{\omega _{0}^{2}+\left|\tilde{\nu }_{k}\right|\zeta +{\tilde{\nu }_{k}}^{2}}
  \end{align}
\end{subequations}
is the mean square of the coordinate $q$, and
\begin{align}
  \tilde{C}_{q}(\omega )&=\frac{1}{2}\coth \left(\frac{\beta \hbar \omega }{2}\right)\frac{1}{\omega _{0}}\frac{\zeta \omega _{0}^{2}\omega }{(\omega _{0}^{2}-\omega ^{2})^{2}+\zeta ^{2}\omega ^{2}}
  \label{eq:analytical-solution}
\end{align}
is the symmetrize-correlation function of the coordinate $q$.

To obtain the thermal equilibrium state numerically, we integrated Eqs.~\eqref{eq:lt-qfpe-d} and \eqref{eq:lt-qse-d} from a temporal initial state to a time sufficiently long that all of the hierarchical elements reached the steady state.
For all of our computations, we fixed the oscillator frequency to $\omega _{0}=400~\mathrm{cm}^{-1}$.
We consider the underdamped ($\zeta =0.1\,\omega _{0}$), critically-damped ($\zeta =2\,\omega _{0}$), and overdamped cases ($\zeta =10\,\omega _{0}$) at the temperature $\beta \hbar \omega _{0}=7.47$ ($T=77~K$), which is in the low-temperature regime.

Because both Eqs.~\eqref{eq:lt-qfpe-d} and \eqref{eq:lt-qse-d} consist of sets of infinitely many differential equations, we need to truncate $\vec{n}$ to carry out numerical calculations.
Here, we adopted the truncation scheme proposed in Refs.~\onlinecite{hartle2013prb, hartle2015prb} with modifications:
The hierarchy is truncated in accordance with the condition that $\vec{n}$ satisfies the relation $\Delta _{\vec{n}}\omega _{0}/\Gamma _{\vec{n}}>\delta _{\mathrm{tol}}$, where $\delta _{\mathrm{tol}}$ is the tolerance of the truncation, with $\Gamma _{\vec{n}}\equiv \sum _{k}^{K}n_{k}\nu _{k}$ and
\begin{align}
  \Delta _{\vec{n}}&\equiv \prod _{k}^{K}\frac{1}{n_{k}!}\left(\frac{\eta _{k}}{\eta _{K}}\right)^{n_{k}}.
\end{align}
For details, see Sec.~\nolink{\ref{sec:s-truncation}} in the Supporting Information.

Numerical calculations were carried out to integrate Eqs.~\eqref{eq:lt-qfpe-d} and \eqref{eq:lt-qse-d} using the fourth-order low-storage Runge-Kutta (LSRK4) method \cite{yan2017cjcp}.
The time step for the LSRK4 method was chosen between $\delta t=0.1\times 10^{-2}~\mathrm{fs}$ and $\delta t=0.5\times 10^{-2}~\mathrm{fs}$.
Uniform meshes were employed to discretize the Wigner function and probability distribution function, and the mesh sizes were set to $N_{q}=64$ and $N_{p}=64$ in the $q$ and $p$ directions, respectively.
The mesh ranges of the Wigner function and probability distribution function in the $q$ direction were chosen between $-4\leq q\leq +4$ and $-12\leq q\leq +12$.
The mesh range of the Wigner function in the $p$ direction was chosen between $-4\leq p\leq +4$ and $-15\leq p\leq +15$.
The finite difference calculations for $q$ and $p$ derivatives in Eqs.~\eqref{eq:lt-qfpe-d} and \eqref{eq:lt-qse-d} were performed using the central difference method with tenth-order accuracy.
For the kinetic term of the Liouvillian in Eq.~\eqref{eq:qm-Liouvillian}, the upwind difference method with ninth-order accuracy was employed for the $q$ derivative \cite{frensley1987prb, kato2013jpcb, sakurai2013jpsj, sakurai2014njp}.
The Moyal products in Eq.~\eqref{eq:qm-potential-term} were evaluated using the discretized convolution representation described in Refs.~\onlinecite{frensley1987prb, frensley1990rmp, kim2007jap} with modifications for multi-state systems (for details of the modifications, see Sec.~\nolink{\ref{sec:s-moyal-product}} of the Supporting Information).
The number of QLT correction terms was chosen from $K=3$ and $4$ for the low-temperature case ($\beta \hbar \omega _{0}=7.47$ $(77~\mathrm{K})$), and $K=2$ was employed for the high-temperature case ($\beta \hbar \omega _{0}=1.92$ $(300~\mathrm{K})$).
The tolerance of the truncation was chosen between $\delta _{\mathrm{tol}}=10^{-4}$ and $\delta _{\mathrm{tol}}=10^{-6}$.
In the case of $\delta _{\mathrm{tol}}=10^{-4}$, the number of total hierarchical elements were $34$ and $14$ for the low-temperature and high-temperature cases, respectively.
The C++ source codes, which allow for the treatment of the phase and coordinate space dynamics with any single-state or multi-state potential forms, are provided as Supporting Information.
The actual numerical integrations for the present calculations were carried out using C++/CUDA codes with cuBLAS and cuFFT libraries to enhance the computational speed with graphics processing unit (GPU).

\begin{figure}
  \centering
  \includegraphics[scale=\SingleColFigScale]{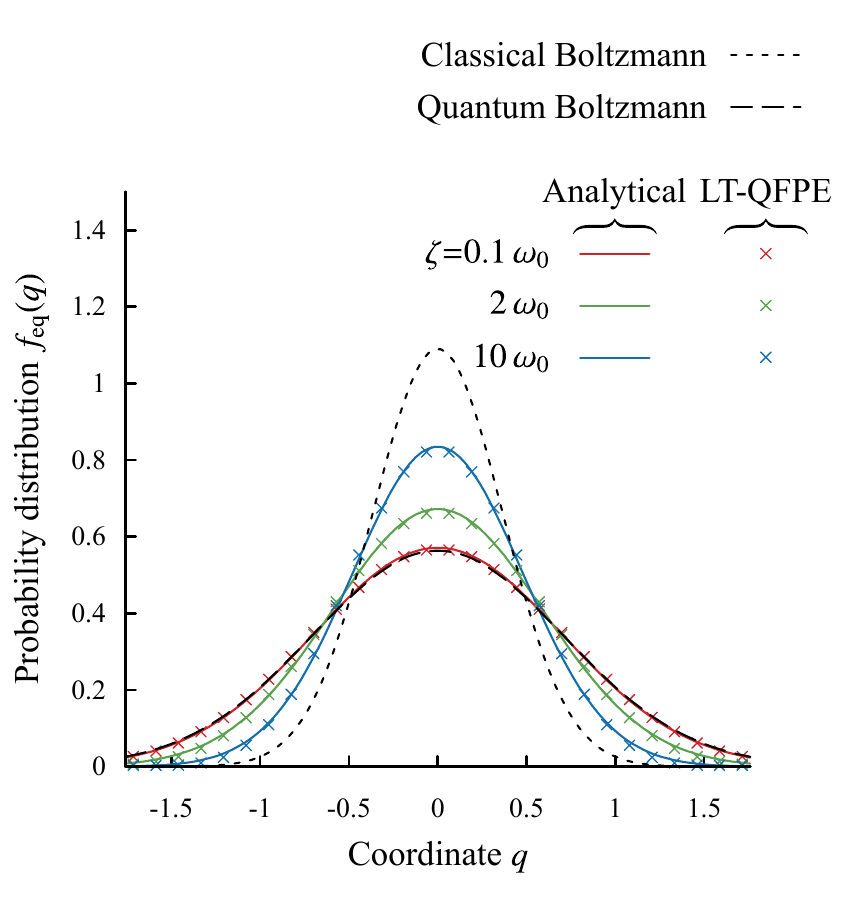}
  \caption{ The equilibrium distributions for a harmonic oscillator in the underdamped ($\zeta =0.1\,\omega _{0}$), critically-damped ($\zeta =2\,\omega _{0}$), and overdamped ($\zeta =10\,\omega _{0}$) cases at low-temperature, $\beta \hbar \omega _{0}=7.47$ $(T=77~\mathrm{K})$.
    The red, green, and blue curves represent the analytically derived solutions, Eq.~\eqref{eq:analytical-distribution}, and the red, green, and blue $\times $ symbols represent the numerical results obtained with the LT-QFPE.
    The classical and quantum equilibrium distributions of the system without a heat bath are also presented as the dotted and dashed curves, respectively. }
  \label{fig:qdistribution}
\end{figure}
\begin{figure}
  \centering
  \includegraphics[scale=\SingleColFigScale]{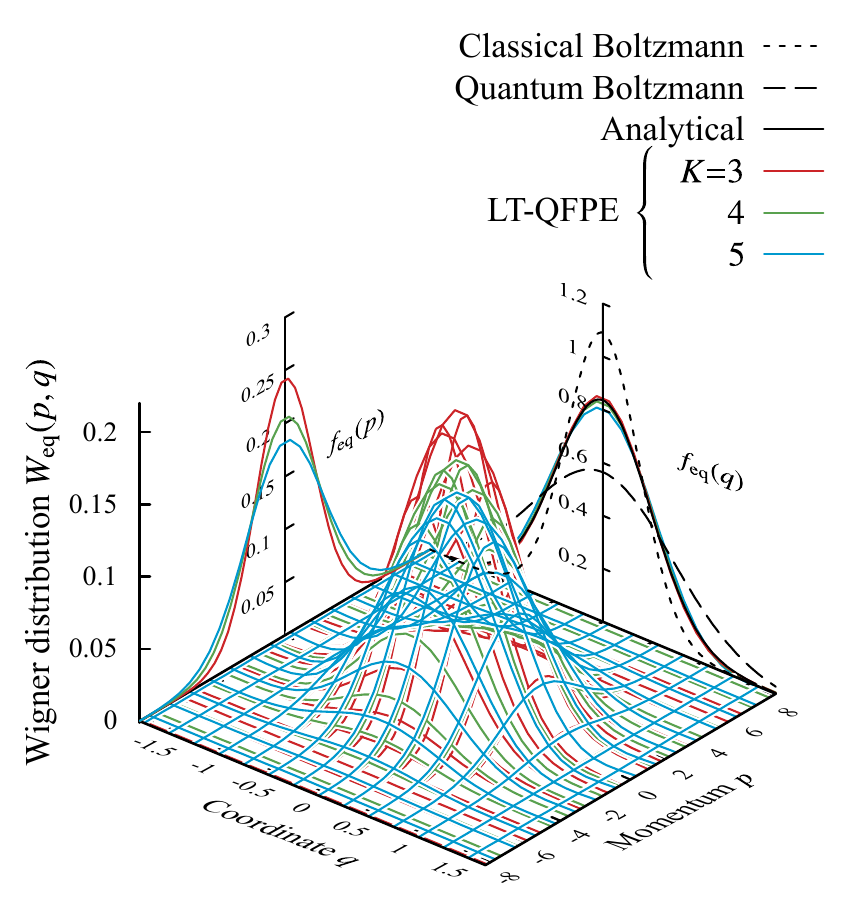}
  \caption{Wigner distribution function for the equilibrium state of a harmonic oscillator in the overdamped case ($\zeta =10\,\omega _{0}$) at $\beta \hbar \omega _{0}=7.47$ $(T=77~\mathrm{K})$ using several values of the number of QLT correction terms ($K=3, 4$ and $5$) .
    The reduced distributions $f(q)$ and $f(p)\equiv \int \!dq\,W(p,q)$ are also displayed. }
  \label{fig:Wdistribution}
\end{figure}
In Fig.~\ref{fig:qdistribution}, we compare the analytic equilibrium distribution with that obtained from the LT-QFPE under several conditions for the system-bath coupling at $\beta \hbar \omega _{0}=7.47$.
The results obtained from the LT-QFPE are overlapped to analytically exact solutions.
In the numerical calculation, the larger number $K$ we used, the more accurate results we had.

In the case of a weak interaction, both the numerical and analytical forms are closer to that for the isolated harmonic oscillator, which is given by Eq.~\eqref{eq:analytical-distribution} with $\langle q^{2}\rangle _{\beta ,0}=(1/2)\coth (\beta \hbar \omega _{0}/2)$.
This expression can also be derived from the Boltzmann summation of the eigenstates.

As mentioned above, the mean-square momentum, $\langle p^{2}\rangle $, diverges in the present Ohmic case \cite{grabert1988pr, weiss2011book}.
This is because high frequency quantum noise destroys the quantum coherence between the ``bra'' and ``ket'' wave functions, which results in the condition $\rho (z,z')=0$ for $z\neq z'$.
As a result, the Wigner distribution function in momentum space, which is the Fourier transform of the quantum coherence $r\equiv z-z'$, is flat in this case (see Sec.~\nolink{\ref{sec:s-uv-divergence}} in the Supporting Information).
However, even in such situations, we can use the Wigner function, because the dynamics of the system are controlled by the low-frequency Matsubara terms or the low-frequency QLT correction terms.

In Fig.~\ref{fig:Wdistribution}, we plot stable solutions for the Wigner distribution function calculated with the LT-QFPE using several values for the number of QLT correction terms.
It is seen that the width of Wigner distributions ($\propto \langle p^{2}\rangle $) increases as the number of QLT correction terms increases, while the $q$ probability distributions converge to on the analytically derived solution.
The Gaussian-like profile in the $p$ direction arises from QLT correction terms for finite $K$.
Although we observe the larger flat distribution for larger $K$, we can still use the Wigner function by ignoring these contribution for the calculation of the non-diverging physical variables.
This is the reason that we can calculate the physical variable using Wigner distribution, while $\langle p^{2}\rangle $ diverges with $K\rightarrow \infty $.

We next study the symmetrized correlation function of the system coordinate, defined as
\begin{align}
  C_{q}(t)&\equiv \mathrm{Tr}\left\{z\mathcal{G}_{\mathrm{tot}}(t)\frac{z+z'}{2}\rho _{\mathrm{tot}}^{\mathrm{eq}}(z,z',\vec{x},\vec{x}\,')\right\},
  \label{eq:symmetrized-correlation}
\end{align}
where $\mathcal{G}_{\mathrm{tot}}(t)$ is Green's function for the total system and $W^{\mathrm{eq}}(p,q,\vec{x},\vec{x})$ is the stationary solution of $\mathcal{G}(t)$.
In the HEOM formalism, the time evolution of the total system, described by $\rho _{\mathrm{tot}}(z,z',\vec{x},\vec{x}\,',t)$, is replaced by that of the hierarchical elements, described by $\rho _{\mathrm{H}}(z,z',t)\equiv \{\rho _{\vec{n}}(z,z',t)\,|\,\vec{n}\in \mathbb{N}^{K}\}$.
It has been found that these yield the same reduced dynamics in the system subspace \cite{tanimura2014jcp, tanimura2015jcp}.
After the Wigner transformation, Eq.~\eqref{eq:symmetrized-correlation} is evaluated as
\begin{align}
  C_{q}(t)&=\int \!dp\,\int \!dq\,q\biggl.\left\{\mathcal{G}_{\mathrm{H}}(t)qW_{\mathrm{H}}^{\mathrm{eq}}(p,q)\right\}\biggr|_{\vec{n}=\vec{0}},
  \label{eq:symmetrized-correlation-heom}
\end{align}  
where $\mathcal{G}_{\mathrm{H}}(t)$ is Green's function evaluated from Eq.~\eqref{eq:lt-qfpe-d} or \eqref{eq:lt-qse-d}, and $W_{\mathrm{H}}^{\mathrm{eq}}(p,q)$ is the stationary solution of $\mathcal{G}_{\mathrm{H}}(t)$.
We define the Fourier transform of Eq.~\eqref{eq:symmetrized-correlation} as
\begin{align}
  C_{q}(\omega )&\equiv \int _{0}^{\infty }\!dt\,C_{q}(t)\cos \omega t.
  \label{eq:ft-symmetrized-correlation}
\end{align}

In the quantum case, Eq.~\eqref{eq:ft-symmetrized-correlation} can be analytically evaluated as Eq.~\eqref{eq:analytical-solution}.
Under the condition $\zeta \gg \omega _{0}$, this function asymptotically approaches \cite{garg1985jcp, tanimura1993pre}
\begin{align}
  \tilde{C}_{q}^{\zeta \gg \omega _{0}}(\omega )&=\frac{1}{2}\coth \left(\frac{\beta \hbar \omega }{2}\right)\frac{1}{\omega _{0}}\frac{\tilde{\gamma }\omega }{\omega ^{2}+\tilde{\gamma }^{2}},
  \label{eq:analytical-solution-overdamped}
\end{align}
where $\tilde{\gamma }\equiv \omega _{0}^{2}/\zeta $.
The classical high-temperature limit of the above result is obtained by replacing $\coth (\beta \hbar \omega /2)$ with $2/\beta \hbar \omega $.
\begin{figure*}
  \centering
  \includegraphics[scale=\DoubleColFigScale]{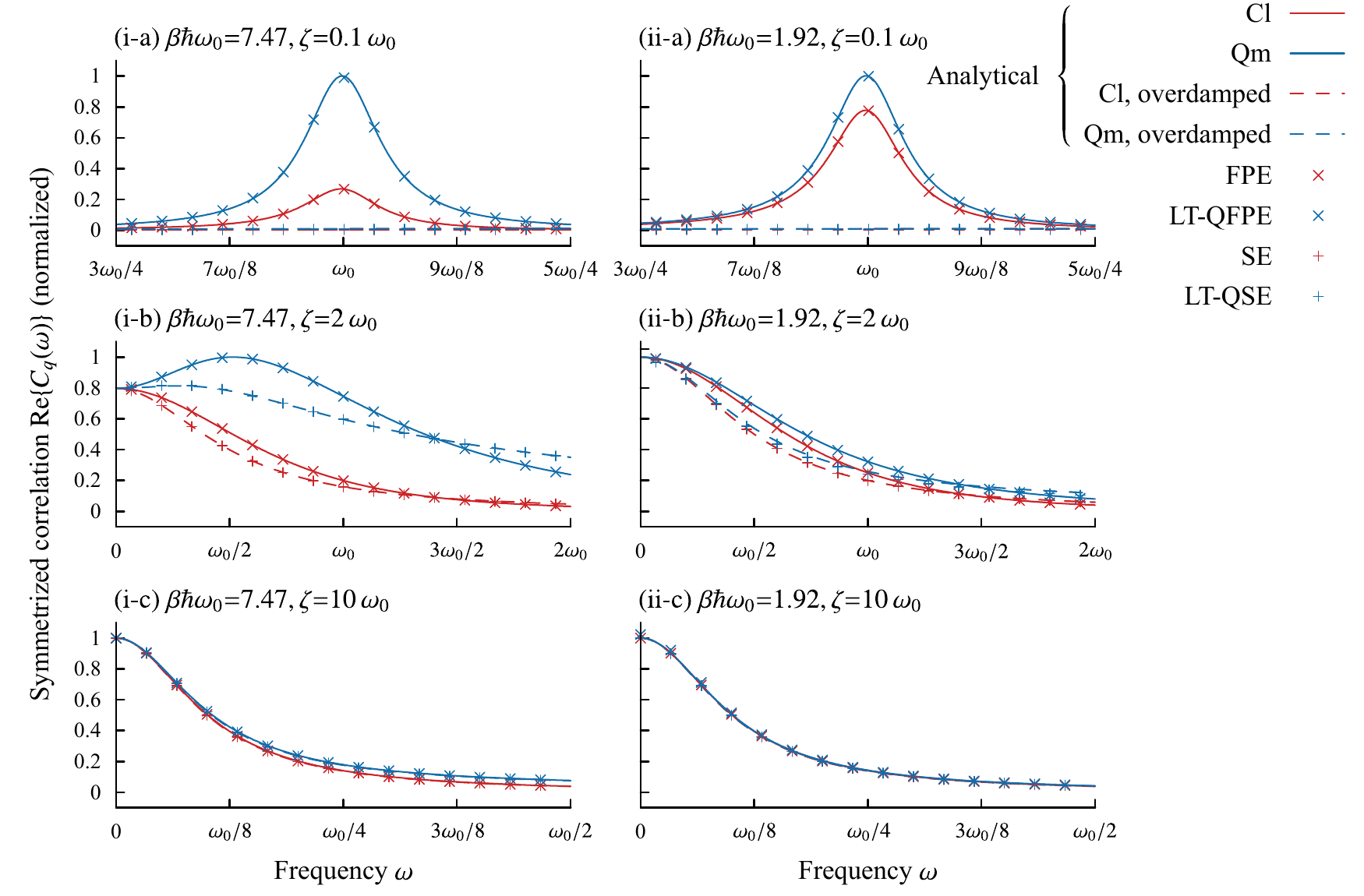}
  \caption{ The symmetrized correlation function given in Eq.~\eqref{eq:symmetrized-correlation} for a harmonic oscillator under (a) underdamped ($\zeta =0.1\,\omega _{0}$), (b) critically-damped ($\zeta =2\,\omega _{0}$), and (c) overdamped conditions ($\zeta =10\,\omega _{0}$) in (i) low-temperature ($\beta \hbar \omega _{0}=7.47$ $(77~\mathrm{K})$) and (ii) high-temperature ($\beta \hbar \omega _{0}=1.92$ $(300~\mathrm{K})$) cases.
    The solid and dotted curves were calculated from Eqs.~\eqref{eq:analytical-solution} and \eqref{eq:analytical-solution-overdamped}, for the classical (red) and quantum (blue) cases, respectively.
    The values plotted here are normalized with respect to the maximum of Eqs.~\eqref{eq:analytical-solution}.
    The LT-QFPE and LT-QSE results are denoted by the blue $\times $ and $+$ symbols, respectively, while the FPE and SE results are denoted by the red $\times $ and $+$ symbols.
    Note that QFPE and the classical Fokker-Planck equation (FPE) are equivalent in the case of a harmonic PES.
  }
  \label{fig:1D}
\end{figure*}
In Fig.~\ref{fig:1D}, we depict the symmetrized-correlation function calculated under several damping conditions in (i) low temperature ($\beta \hbar \omega _{0}=7.47$) and (ii) high temperature ($\beta \hbar \omega _{0}=1.92)$ cases.
As seen there, the numerical results obtained from Eqs.~\eqref{eq:lt-qfpe-d} and \eqref{eq:lt-qse-d} are close to the exact analytical solutions.
Generally, in the HEOM formalism, we are able to obtain as accurate results as we need by employing larger hierarchical space (i.e.~by increasing $K$ and by decreasing $\delta _{\mathrm{tol}}$).
It should be noted that, although the LT-QSE well predicts the quantum dynamics under strong friction, the stable solution of the LT-QSE depend upon the number of QLT correction terms.
As shown in Appendix~\ref{sec:appendx-qse}, because the analytical solution for the overdamped case, Eq.~\eqref{eq:analytical-solution-overdamped}, also diverges under the infinite summation of the Matsubara frequencies, careful verification is important when we use the LT-QSE to calculate observable strongly depend on the equilibrium distribution.

As shown in this section, the LT-QFPE and LT-QSE can describe accurate dynamics with a properly truncated hierarchical space.
This finding is important for numerical calculations.
As shown in Appendix \ref{sec:appendx-langevin}, the LT-QFPE and LT-QSE are equivalent to the Langevin expressions for a single harmonic potential case.
As analytically calculated symmetrized correlation functions from the Langevin equations indicate, we can obtain Eqs.~\eqref{eq:analytical-solution} and \eqref{eq:analytical-solution-overdamped} from the present approach, under $\delta _{\mathrm{tol}}\rightarrow 0$ and $K\rightarrow \infty $.
This result demonstrates the reliability of the LT-QFPE and LT-QSE theories.

\subsection{Multi-State Case: Double-Well PESs with Gaussian Adiabatic Coupling }
Next, we present our numerical results for multi-state systems.
For convenience, we describe our system using adiabatic electronic states, $|\Phi _{a}(z)\rangle $.
In the following, $a$, $b$ and $c$ refer to adiabatic electronic states, and $j$, $k$ and $l$ refer to diabatic electronic states.

\subsubsection{Adiabatic and diabatic bases}
The $a$th adiabatic electronic state is an eigenfunction of the time-independent Schr\"odinger equation, and thus we have
\begin{align}
  \hat{U}(z)|\Phi _{a}(z)\rangle &=U^{\mathrm{a}}_{a}(z)|\Phi _{a}(z)\rangle ,
  \label{eq:eigen}
\end{align}
where $\hat{U}(z)\equiv \sum _{j,k}|j\rangle U^{\mathrm{d}}_{jk}(z)\langle k|$, and $U^{\mathrm{a}}_{a}(z)$ is the $a$th adiabatic Born-Oppenheimer (BO) PES.
The diabatic and adiabatic states are related through the transformation matrix given by
\begin{align}
  Z_{ja}(z)\equiv \langle j|\Phi _{a}(z)\rangle 
  \label{eq:transformation-matrix}.
\end{align}
We introduce the unitary matrix $\bm{Z}(z)$ defined as $\{\bm{Z}(z)\}_{ja}=Z_{ja}(z)$, which satisfies the relation $\bm{Z}(z)^{\dagger }\bm{Z}(z)=\bm{Z}(z)\bm{Z}(z)^{\dagger }=\bm{1}$.
Then, Eq.~\eqref{eq:eigen} can be expressed in diagonal matrix form as
\begin{align}
  \bm{Z}(z)^{\dagger }\bm{U}^{\mathrm{d}}(z)\bm{Z}(z)&=\bm{U}^{\mathrm{a}}(z)
  \label{eq:diagonalization},
\end{align}
where $\{\bm{U}_{\mathrm{a}}(z)\}_{ab}\equiv \delta _{ab}U_{\mathrm{a}}^{a}(z)$, and $\delta _{ab}$ is the Kronecker delta.

\begin{subequations}
  In adiabatic representations of kinetic equations, non-adiabatic couplings between adiabatic states are characterized by the non-adiabatic coupling matrix, $\bm{d}(z)$, expressed in terms of the first-order derivative of the coordinate as \cite{stock2005acp, may2008book}
  \begin{align}
    \{\bm{d}(z)\}_{ab}=d_{ab}(z)&\equiv \langle \Phi _{a}(z)|\frac{\partial }{\partial z}|\Phi _{b}(z)\rangle .
  \end{align}
  The non-adiabatic coupling matrix, $\bm{d}$, is skew-Hermitian (i.e.~$\bm{d}^{\dagger }={-}\bm{d}$).
  This matrix can also be expressed in terms of $\bm{Z}(z)$ as
  \begin{align}
    \bm{d}(z)&=\bm{Z}(z)^{\dagger }\frac{\partial }{\partial z}\bm{Z}(z), 
  \end{align}
\end{subequations}
and therefore we have
\begin{align}
  \bm{Z}(z)&=\bm{Z}(-\infty )\mathop{\exp }_{\rightarrow }\left(\int _{-\infty }^{z}\!dz'\,\bm{d}(z')\right),
  \label{eq:d-Z-transform}
\end{align}
where $\mathop{\exp }_{\rightarrow }$ is the ordered exponential in coordinate space.
Thus, the transformation matrix, Eq.\eqref{eq:transformation-matrix}, can be constructed from $\bm{d}(z)$, and hence the diabatic PESs can be obtained from the adiabatic PESs using the inverse of the transformation in Eq.~\eqref{eq:diagonalization}.

\begin{subequations}
  If necessary, we can introduce the non-adiabatic coupling matrix of the second-order, defined as
  \begin{align}
    \{\bm{h}(z)\}_{ab}=h_{ab}(z)&\equiv \langle \Phi _{a}(z)|\frac{\partial ^{2}}{\partial z^{2}}|\Phi _{b}(z)\rangle ,
  \end{align}
  which can be constructed from $\bm{d}(z)$ as
  \begin{align}
    \bm{h}(z)&=\bm{Z}(z)^{\dagger }\frac{\partial ^{2}}{\partial z^{2}}\bm{Z}(z)=\frac{\partial \bm{d}(z)}{\partial z}+\bm{d}(z)^{2}.
  \end{align}
\end{subequations}

Next, we introduce the reduced density matrix in the adiabatic representation, defined as
\begin{align}
  \rho _{ab}^{\mathrm{a}}(z,z',t)&\equiv \langle \Psi _{a}(z)|\hat{\rho }(z,z',t)|\Psi _{b}(z')\rangle ,
\end{align}
where the diagonal element $\rho _{aa}^{\mathrm{a}}(z,z,t)$ and the off-diagonal element $\rho _{ab}^{\mathrm{a}}(z,z',t)$ $(a{\neq }b)$ represent the population of $|\Phi _{a}(z)\rangle $ and the coherence between $|\Phi _{a}(z)\rangle $ and $|\Phi _{b}(z')\rangle $, respectively.
The adiabatic representation of the density matrix, $\{\bm{\rho }^{\mathrm{a}}(z,z',t)\}_{ab}=\rho _{ab}^{\mathrm{a}}(z,z',t)$, can be obtained from $\bm{\rho }^{\mathrm{d}}(z,z',t)$ through application of the transformation matrix $\bm{Z}(z)$ as
\begin{align}
  \bm{\rho }^{\mathrm{a}}(z,z',t)&=\bm{Z}\left(z\right)^{\dagger }\bm{\rho }^{\mathrm{d}}(z,z',t)\bm{Z}\left(z'\right).
  \label{eq:adiabatic-rep}
\end{align}
This representation is related to the Wigner representation as
\begin{align}
  \bm{W}^{\mathrm{a}}(p,q,t)&\equiv \frac{1}{2\pi }\int \!dr\,e^{-ipr}\bm{\rho }^{\mathrm{a}}\left(q+\frac{r}{2},q-\frac{r}{2}\right).
  \label{eq:adiabtic-transform}
\end{align}
Although we can construct the equations of motion for $\bm{W}^{\mathrm{a}}(p,q,t)$ directly, the numerical integrations are complicated, because the number of terms that include the Moyal product becomes large (see Appendix~\ref{eq:app-adiabatic-rep}).
For this reason, we integrate the equations of motion in the diabatic representation.
After obtained the numerical results, we convert these to the adiabatic representation.

\subsubsection{Tilted double-well model}
\begin{subequations}
  As a schematic model for IC in a photoisomerization process, we adopt the following tilted double-well adiabatic ground BO PES:
  \begin{align}
    U_{g}^{\mathrm{a}}(q)&=\frac{\hbar \omega _{0}}{2L_{0}^{2}}q^{2}\left(q^{2}-\frac{L_{0}^{2}}{2}\right)+\frac{\Delta E}{L_{0}}q.
    \label{eq:U_g}
  \end{align}
  Here, $L_{0}$ and $\Delta E$ are the displacement between the wells and the difference between their energies, respectively.
  We use the following harmonic adiabatic excited BO PES:
  \begin{align}
    U_{e}^{\mathrm{a}}(q)&=\frac{\hbar \omega _{e}^{2}}{2\omega _{0}}(q-q^{\dagger })^{2}+U_{g}^{\mathrm{a}}(q^{\dagger })+E_{\mathrm{gap}}^{e-g}.
    \label{eq:U_e}
  \end{align}
  Here, $\omega _{e}$, $q^{\dagger }$ and $E_{\mathrm{gap}^{e-g}}$ are the vibrational characteristic frequency in the excited state, the position of the crossing region, and the energy gap between the ground and excited BO PES in the crossing region, respectively.
\end{subequations}

We assume that the non-adiabatic coupling has the Gaussian form
\begin{align}
  d_{eg}(q)&=-d_{ge}(q)=\sqrt {\mathstrut \frac{\pi }{8\sigma ^{\dagger 2}}}e^{-(q-q^{\dagger })^{2}/2\sigma ^{\dagger 2}},
  \label{eq:d_eg}
\end{align}
and $d_{gg}(q)=d_{ee}(q)=0$, where $\sigma ^{\dagger }$ is the width of the crossing region.
The integral of $d_{eg}(q)$ is given by
\begin{align}
  \begin{split}
    D_{eg}(q)&\equiv \int _{-\infty }^{q}\!dq'\,d_{eg}(q')=\frac{\pi }{4}\mathrm{erfc}\left(-\frac{q-q^{\dagger }}{\sqrt {\mathstrut 2}\sigma ^{\dagger }}\right),
  \end{split}
\end{align}
where $\mathrm{erfc(x)}\equiv 1-\mathrm{erf}(x)$ is the complementary error function.
Because $d_{eg}(q)$ is normalized with respect to $D_{eg}(\infty )=\pi /2$, the adiabatic bases are exchanged with the change of position from $z=-\infty $ to $z=+\infty $.

Although we can construct the MS-LT-QFPE and their variant equations in the adiabatic representation (see Appendix~\ref{eq:app-adiabatic-rep}), in the present study, we performed the numerical calculation using the diabatic representation, because in this case, the equations are simpler and easier to solve.
Then, after we obtained the numerical results, we carried out the inverse transformation in order to convert these to the adiabatic representation.
We employ the diabatic basis defined as $|0\rangle \equiv |\Phi _{g}(-\infty )\rangle $ and $|1\rangle \equiv |\Phi _{e}(-\infty )\rangle $.
Then Eq.~\eqref{eq:d-Z-transform} is solved as
\begin{align}
  \begin{split}
    \bm{Z}(q)
    = \begin{pmatrix}
      \quad \cos D_{eg}(q) & -\sin D_{eg}(q)\\
      +\sin D_{eg}(q) & \quad \cos D_{eg}(q)
    \end{pmatrix},
  \end{split}
\end{align}
and the diabatic PESs and coupling are given by
\begin{align}
  \left\{
  \begin{aligned}
    U_{00}^{\mathrm{d}}(q)=\frac{U_{e}^{\mathrm{a}}(q)+U_{g}^{\mathrm{a}}(q)}{2}&-\cos \bigl(2D_{eg}(q)\bigr)\frac{U_{e}^{\mathrm{a}}(q)-U_{g}^{\mathrm{a}}(q)}{2}\\
    U_{11}^{\mathrm{d}}(q)=\frac{U_{e}^{\mathrm{a}}(q)+U_{g}^{\mathrm{a}}(q)}{2}&+\cos \bigl(2D_{eg}(q)\bigr)\frac{U_{e}^{\mathrm{a}}(q)-U_{g}^{\mathrm{a}}(q)}{2}\\
    U_{10}^{\mathrm{d}}(q)=U_{01}^{\mathrm{d}}(q)=&-\sin \bigl(2D_{eg}(q)\bigr)\frac{U_{e}^{\mathrm{a}}(q)-U_{g}^{\mathrm{a}}(q)}{2}.\\
  \end{aligned}
  \right.
  \label{eq:diabatic-PESs-01}
\end{align}

\begin{table}
  \centering
  \caption{ The parameter values for the numerical tests. }
  \begin{tabular}{c@{~}r@{~}l@{\quad}c@{~}r@{~}l@{\quad}c@{~}r@{~}l}
    \hline
    \hline
    Symbol & \multicolumn{2}{c}{Value} &
    Symbol & \multicolumn{2}{c}{Value} &
    Symbol & \multicolumn{2}{c}{Value} \\
    \hline
    \hline
    $\omega _{\mathrm{0}}$ & $400$&$\mathrm{cm}^{-1}$ &
    $\omega _{e}$ & $\omega _{0}$& &
    $\omega _{f}$ & $1.5\,\omega _{0}$& \\
    $\Delta E$ & $2,000$&$\mathrm{cm}^{-1}$ &
    $E_{\mathrm{gap}}^{e-g}$ & $1,000$&$\mathrm{cm}^{-1}$ &
    $E_{\mathrm{gap}}^{f-e}$ & $5,000$&$\mathrm{cm}^{-1}$ \\
    $L_{0}$ & $10$& &
    $q^{\dagger }$ & $1$& &
    $\sigma ^{\dagger }$ & $1$& \\
    \hline
    \hline
  \end{tabular}
  \label{tab:parameters}
\end{table}
\begin{figure}
  \centering
  \includegraphics[scale=\SingleColFigScale]{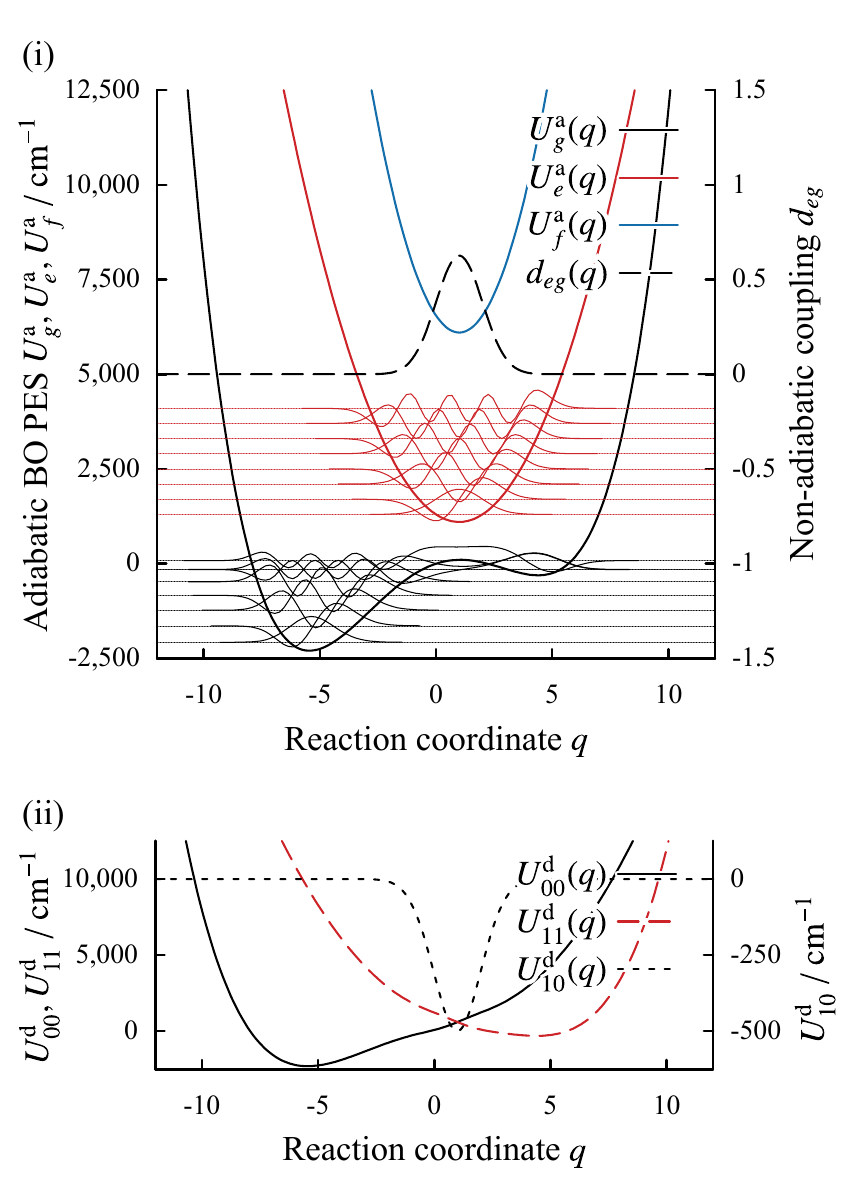}
  \caption{ (i) The adiabatic ground BO PES (black solid curve), excited BO PES (red solid curve), and non-adiabatic coupling (black dashed curve) given by Eqs.~\eqref{eq:U_g}, \eqref{eq:U_e} and \eqref{eq:d_eg} are depicted for the parameter values given in Table~\ref{tab:parameters}.
    The blue solid curve represents the second excited BO PES.
    The first eight vibrational eigenfunctions for the ground and first excited BO PESs are also plotted.
    (ii) The diabatic $00$ PES (black solid curve), diabatic $11$ PES (red dashed curve), and the diabatic $10$ coupling (black dotted curve) for the case of (i). }
  \label{fig:PES}
\end{figure}
Figure~\ref{fig:PES} presents the adiabatic BO PESs and diabatic PESs for the parameter values listed in Table~\ref{tab:parameters}.

We set the initial distribution as
\begin{align}
  W_{ee}^{\mathrm{a}}(p,q,0)=\frac{1}{\mathcal{Z}}
  e^{-\tanh (\beta \hbar \omega _{0}/2)\left[p^{2}+(q-q_{\mathrm{i}})^{2}\right]},
  \label{eq:initial}
\end{align}
and as $W_{gg}^{\mathrm{a}}(p,q,0)=W_{eg}^{\mathrm{a}}(p,q,0)=W_{ge}^{\mathrm{a}}(p,q,0)=0$ in Eq.~\eqref{eq:initial}, where $q_{\mathrm{i}}=-L_{0}/2$ and $\mathcal{Z}$ is the partition function.
This is the Wigner transformation of the Boltzmann distribution for the harmonic oscillator centered at $q=q_{\mathrm{i}}$.
In this demonstration, we ignore the initial correlation at $t=0$.

We performed the numerical calculations to integrate equations of motion using the finite difference method with mesh sizes $N_{q}=256$ and $N_{p}=64$ and mesh ranges $-12\leq q\leq +12$ and $-12\leq p\leq +12$.
  The other calculation conditions were the same as in Sec.~\ref{sec:brownian}.
For comparison, we display the results calculated using the fewest switch surface hopping (FSSH) \cite{tully1990jcp, hammes1994jcp} and Ehrenfest methods \cite{stock2005acp, may2008book} with a classical Markovian Langevin force under the same conditions.
In both methods, the adiabatic electronic basis was employed.
In the Ehrenfest methods, the state of the system is described using the electronic density matrix (or the electronic wavefunction), $\bm{\rho }^{\mathrm{el}}(t)$, and a trajectory, $\{p(t), q(t)\}$, which is determined following the mean-field force calculated from the electronic PESs (i.e.~averaged force from the populations in $\bm{\rho }^{\mathrm{el}}(t)$).
In the FSSH methods, the state is also described using $\bm{\rho }^{\mathrm{el}}(t)$, but its trajectory, $\{p(t), q(t)\}$, now follows the force calculated from the active PES, $U_{\lambda (t)}^{\mathrm{a}}(q)$, where $\lambda (t)$ is a randomly changing index (i.e.~$g$ or $e$) in time whose hopping rate is calculated using $\bm{\rho }^{\mathrm{el}}(t)$.
In each case, the time evolution of the particle trajectories under the Langevin force was calculated using the Vanden-Eijnden-Ciccotti methods \cite{tuckerman2010book}.
The time evolution of $\bm{\rho }^{\mathrm{el}}(t)$ was calculated by the numerical integration of the Schr\"odinger equation (NISE) method \cite{torii2006jpca, jansen2006jpcb}, in which the coefficients of the time evolution operator of the electronic density matrix were held constant during each time step evaluation.
We used $\delta t=0.1~\mathrm{fs}$ as the time step for the integrations using the FSSH and Ehrenfest methods, and we employed $N\simeq 10,000$ trajectories for the average calculations.

\subsubsection{Population dynamics}
\begin{figure*}
  \centering
  \includegraphics[scale=\DoubleColFigScale]{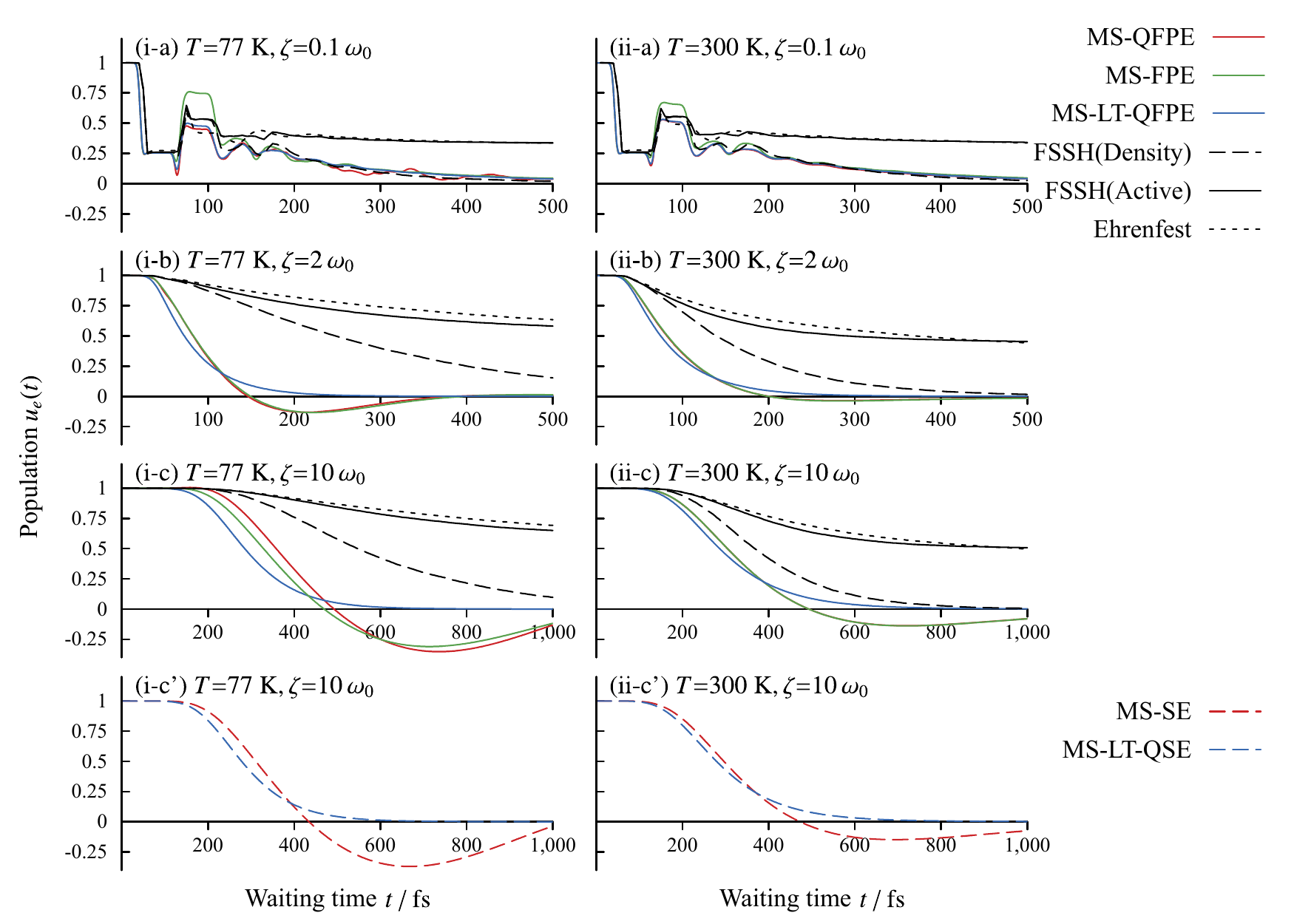}
  \caption{ The time evolution of the excited population in the adiabatic state $|\Phi _{e}^{\mathrm{a}}\rangle $ under (a) underdamped ($\zeta =0.1\,\omega _{0}$), (b) critically-damped ($\zeta =2\,\omega _{0}$), and (c) overdamped conditions ($\zeta =10\,\omega _{0}$) in the (i) low-temperature ($77~\mathrm{K}$) and (ii) high-temperature ($300~\mathrm{K}$) cases, respectively.
    The red and blue solid curves were calculated using the MS-QFPE and MS-LT-QFPE, respectively.
    The green solid curves represent the MS-FPE results.
    These were derived using the first-order Moyal truncated Liouvillian, which is equivalent to the mixed quantum-classical Liouvillian \cite{kapral1999jcp, ikeda2018cp}.
    The black solid and dotted curves are the ensemble averages of the excited state population of the electronic density matrix, $\rho ^{\mathrm{el}}_{ee}(t)$, calculated in the FSSH and Ehrenfest cases, respectively.
    The black dashed curves are the excited population, $\bar{u}_{\lambda (t){=}e}(t)$, in the FSSH (active) case.
    Figs.~(i-c') and (ii-c') correspond to the Smoluchowski limit, which can be evaluated from from Eq.~\eqref{eq:lt-qse-d}. }
  \label{fig:population}
\end{figure*}
In Fig.~\eqref{fig:population}, we present plots of the excited population in the adiabatic representation defined as,
\begin{align}
  u_{e}(t)&\equiv \int \!dp\,\int \!dq\,W_{ee}^{\mathrm{a}}(p,q,t).
\end{align}
In the FSSH and Ehrenfest cases, we defined $u_{e}(t)$ as an ensemble average of $\rho ^{\mathrm{el}}_{ee}(t)$ for all trajectories. 
  In the FSSH (active) case, we further introduced the excited population, $\bar{u}_{\lambda (t){=}e}(t)$, calculated as an ensemble average of the population with the active PES index $\lambda (t)=e$ for all trajectories.

As demonstrated in Sec.~\ref{sec:brownian}, the MS-LT-QFPE is numerically accurate, and therefore the calculated results obtained with these equations can be used as a reference to verify other results from different formalisms.
We now discuss the numerical results.
A wavepacket centered at $q=q_{\mathrm{i}}$ in the excited state as the initial state.
The wavepacket then moves in the direction of the crossing region, i.e.~near $q=q^{\dagger }$.
Then the population of $|\Phi _{e}(q)\rangle $ decreases due to the non-adiabatic transition.
In the case that $\zeta $ is sufficiently small, the de-excited wavepacket moves among the double minima of the adiabatic ground state PES while maintaining a large kinetic energy, and it traverses the crossing region repeatedly through non-adiabatic transitions.
As a result, oscillatory behavior is observed in Figs.~\eqref{fig:population}(i-a) and \eqref{fig:population}(ii-a).
As the coupling constant $\zeta $ increases, the wavepacket motion becomes slower, and the population of $|\Phi _{e}(q)\rangle $ decays more slowly.
While classical treatments of the heat bath produce pathological negative populations (red and green curves), in particular in the low temperature case, the presently investigated equations of motion, Eqs.~\eqref{eq:lt-qfpe-d} and \eqref{eq:lt-qse-d}, accurately describe the population dynamics (the blue lines).
This demonstrates the importance of the QLT correction terms.

While the results obtained with the FSSH and Ehrenfest methods do not exhibit negative populations because of the assumptions for a decomposition of the distribution function into the trajectories, they do differ significantly from both the quantum (MS-LT-QFPE) and semi-classical (MS-FPE) results, particularly in the low temperature case.
Because the FSSH and Ehrenfest methods were originally developed to study isolated systems, which have a few degrees of freedom, these methods may not be proper for calculations of systems in dissipative conditions that are essentially many-body problem when we include the bath degrees of freedom.
It should be noted that, in the FSSH results, $u_{e}(t)$ and $\bar{u}_{\lambda (t){=}e}(t)$ disagrees, while both correspond the electronic excited state.
The similar disagreement is reported in Ref.~\onlinecite{tempelaar2013jcp}.

Note that, as the coupling constant $\zeta $ increases, the results obtained from the quantum Liouvillian (red curves) and semi-classical Liouvillian (green curves) treatments of the PES become similar.
This is because the higher-order differential operators in the Moyal products vanish in the overdamped limit, as shown in the construction of the LT-QSE (see Sec.~\nolink{\ref{sec:s-lt-qse}} in Supporting Information).
Moreover, when the energy gap between the ground and excited states in the crossing region is small, the negative population of the semi-classical results is suppressed.
This is because, the characteristic frequency of the electronic transition dynamics becomes small in such situation, and therefore the high-temperature approximation of the bath works well, as the authors' previous investigation for conical intersection problem \cite{ikeda2018cp}.

\subsection{Transient Absorption Spectra}
The presently investigated formalisms are capable of calculating non-linear response functions.
By calculating the transient absorption spectrum, here we demonstrate the importance of the QLT correction terms for nonlinear optical spectra.
In order to include an excited state absorption (ESA) process, we add a second excited state, $|\Phi _{f}(q)\rangle $, with the PES
\begin{align}
  U_{f}^{\mathrm{a}}(q)&=\frac{\hbar \omega _{f}^{2}}{2\omega _{0}}(q-q^{\dagger })^{2}+U_{e}^{\mathrm{a}}(q^{\dagger })+E_{\mathrm{gap}}^{e-f},
\end{align}
to the present model.
Because we assume $d_{fg}(q)=d_{fe}(q)=0$, a spontaneous transition between $|\Phi _{f}(q)\rangle $ and the subspace $\{|\Phi _{g}(q)\rangle , |\Phi _{e}(q)\rangle \}$ is prohibited.
Thus, only optically stimulated transitions between $|\Phi _{g}(q)\rangle $ and $|\Phi _{e}(q)\rangle $ occur.
The diabatic PESs of this three-state system are those given in Eq.~\eqref{eq:diabatic-PESs-01}, along with $U_{22}(q)=U_{f}(q)$, $U_{20}(q)=U_{21}(q)=0$.

The transient absorption (TA) spectrum from the initial state Eq.~\eqref{eq:initial} is given by \cite{ikeda2017jcp}
\begin{align}
  I^{\mathrm{TA}}(\omega ,t)&\equiv \omega \mathrm{Im}\int _{0}^{\infty }\!d\tau \,e^{i\omega \tau }R^{\mathrm{TA}}(\tau ,t),
  \label{eq:ta-spectrum}
\end{align}
\begin{widetext}
  where
  \begin{align}
    \begin{split}
      R^{\mathrm{TA}}(\tau ,t)&\equiv \mathrm{Tr}_{\mathrm{S}}\biggl\{\bm{\mu }^{\mathrm{d}}(q)\mathrm{Tr}_{\mathrm{B}}\Bigl\{\mathcal{G}_{\mathrm{tot}}(\tau )\frac{i}{\hbar }\bm{\mu }^{\mathrm{d}}(q)^{\times }\mathcal{G}_{\mathrm{tot}}(t)\left(\bm{W}^{\mathrm{d}}(p,q,0)\otimes \rho _{\mathrm{B}}^{\mathrm{eq}}(\vec{x},\vec{x}')\right)\Bigr\}\biggr\}
    \end{split}
    \label{eq:ta-response}
  \end{align}
is the response function of the TA and $\{\bm{\mu }^{\mathrm{d}}(q)\}_{jk}=\langle j|\hat{\mu }(q)|k\rangle $ is the dipole operator in the matrix representation.
Here, $A^{\times }B\equiv AB-BA$ is the commutator, and 
\begin{align}
  \mathrm{Tr}_{\mathrm{S}}\{\dots \}&\equiv \int \!dp\,\int \!dq\,\sum _{j}\{\dots \}_{jj}
\end{align}
is the trace over the system subspace.
In the HEOM formalism, Eq.~\eqref{eq:ta-response} becomes
  \begin{align}
    \begin{split}
      R^{\mathrm{TA}}(\tau ,t)&=\mathrm{Tr}_{\mathrm{S}}\biggl\{\bm{\mu }^{\mathrm{d}}(q)\Bigl.\Bigl\{\mathcal{G}_{\mathrm{H}}(\tau )\frac{i}{\hbar }\bm{\mu }^{\mathrm{d}}(q)^{\times }\mathcal{G}_{\mathrm{H}}(t)
      \bm{W}_{\mathrm{H}}^{\mathrm{d}}(p,q,0)\Bigr\}\Bigr|_{\vec{n}=\vec{0}}\biggr\},
    \end{split}
    \label{eq:ta-response-heom}
  \end{align}
\end{widetext}
where $\bm{W}_{\mathrm{H}}^{\mathrm{d}}(p,q,0)\equiv \{\bm{W}_{\vec{n}}^{\mathrm{d}}(z,z',t)\,|\,\vec{n}\in \mathbb{N}^{K}\}$ is initialized as $\bm{W}_{\vec{n}}^{\mathrm{d}}(p,q,0)=\bm{W}^{\mathrm{d}}(p,q,0)$ for $\vec{n}=\vec{0}$ and $\bm{W}_{\vec{n}}^{\mathrm{d}}(p,q,0)=0$ otherwise.

Hereafter, for $\hat{\mu }(q)$, we assume the form $\hat{\mu }(q)=|\Phi _{e}(q)\rangle \langle \Phi _{g}(q)|+|\Phi _{f}(q)\rangle \langle \Phi _{e}(q)|+\mathrm{c.~c}$.
While $\hat{\mu }(q)$ may induce non-vertical transitions among adiabatic electronic states, here we consider the vertical transition only for simplicity (i.e.~the laser interaction is described by the commutator, $\bm{\mu }^{\mathrm{a}}(p,q)^{\times }\bm{W}^{\mathrm{a}}(p,q)$ among the electronic states).
Note that because the distribution, Eq.~\eqref{eq:initial}, is set in the excited state at the initial time, ground state bleaching (GSB) is not observed.

\begin{figure*}
  \centering
  \includegraphics[scale=\DoubleColFigScale]{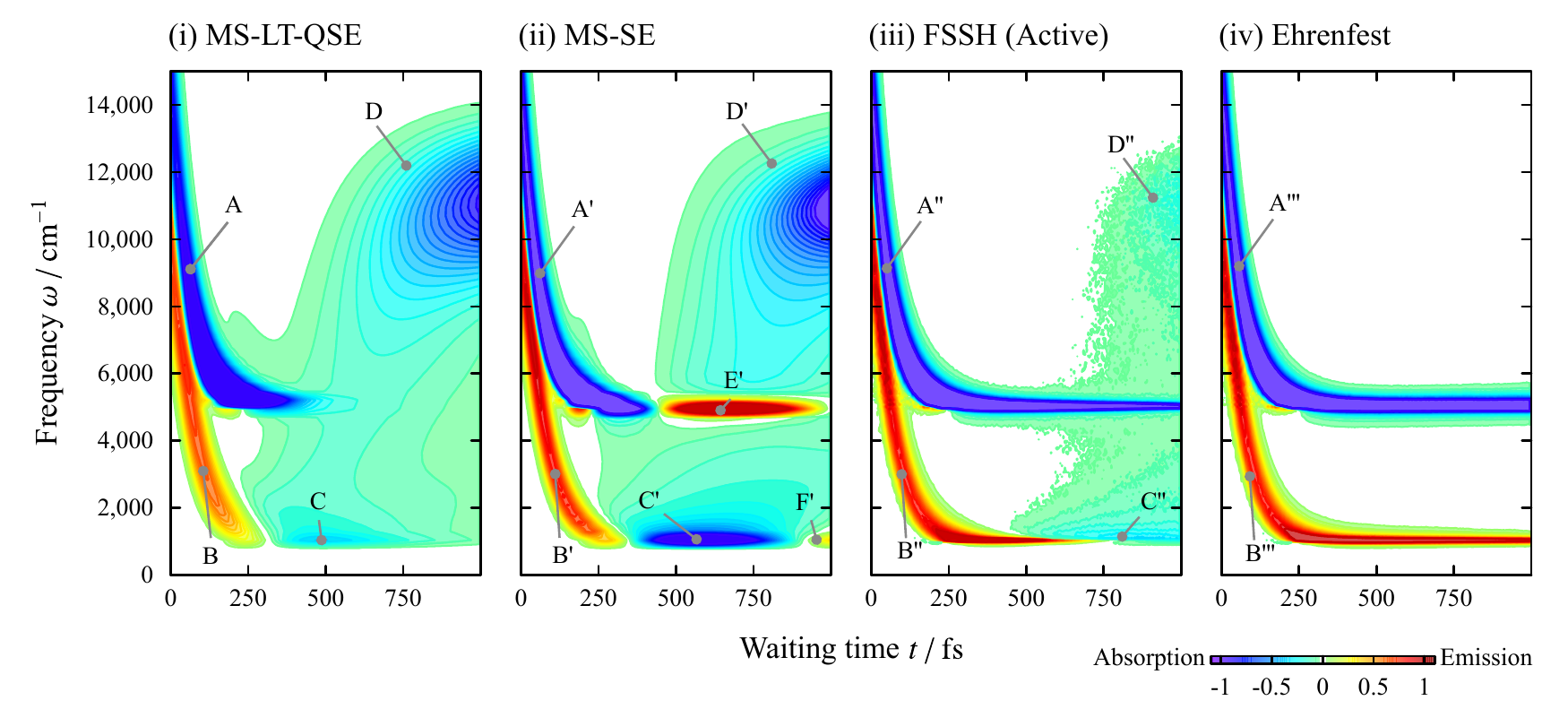}
  \caption{
    Transient absorption spectra, $I^{\mathrm{TA}}(\omega ,t)$, calculated in the overdamped low-temperature case ( $\zeta =10\,\omega _{0}$ and $T=77~\mathrm{K}$).
    The results were obtained using the (i) MS-LT-QSE, (ii) MS-SE, and (iii) FSSH, and (iv) Ehrenfest methods, respectively.
    In the FSSH method, the effects of the non-adiabatic transition dynamics during waiting time was taken into account using the ensemble average of the active PES index, $\bar{u}_{\lambda (t){=}e}(t)$.
    The red and blue regions represent emission and absorption, respectively.
    The values are normalized with respect to the maximum of the MS-LT-QSE calculation at $t=0~\mathrm{fs}$.
  }
  \label{fig:transient-absorption}
\end{figure*}
Figure~\ref{fig:transient-absorption} displays the calculated results in the overdamped case at $T=77~\mathrm{K}$.
Here, the parameter values are the same as in Fig.~\ref{fig:population}(i-c').
For comparison, we display the TA spectra obtained using the FSSH and Ehrenfest methods with a classical Markovian Langevin force under the same conditions.
As shown in Fig.~\ref{fig:population}, non-adiabatic transitions in the FSSH and Ehrenfest calculations are slow.
This is reflected by very sharp peaks in the TA spectra.
To avoid numerical errors in the computation of the Fourier transform due to these sharp peaks, we employed an exponential filter, $\exp (-\tau /\tau _{\mathrm{d}})$, with decay time constant $\tau _{\mathrm{d}}=200~\mathrm{fs}$ in the all calculations.
The procedure of calculating non-linear response functions using the FSSH and Ehrenfest methods following that for the two-dimensional electronic spectra given in Ref.~\onlinecite{tempelaar2013jcp} is presented in Sec.~\nolink{\ref{sec:s-trajectory-based-methods}} of the Supporting Information.

To carry out FSSH calculations, the effects of the non-adiabatic transition dynamics during waiting time was taken into account using the ensemble average of the active PES index, $\bar{u}_{\lambda (t){=}e}(t)$ (i.e.~using the auxiliary wavefunction given in Ref.~\onlinecite{tempelaar2013jcp}).

First, we discuss the TA spectrum plotted in Fig.~\ref{fig:transient-absorption}(i).
At $t=0$, ESA and stimulated emission (SE) are observed at frequencies corresponding to the Frank-Condon point, $U_{f}(q_{\mathrm{i}})-U_{e}(q_{\mathrm{i}})=14,000~\mathrm{cm}^{-1}$ and $U_{e}(q_{\mathrm{i}})-U_{g}(q_{\mathrm{i}})=10,552~\mathrm{cm}^{-1}$, respectively.
The ESA signals is labeled ``$\mathrm{A}$'' and the SE signal is labeled  ``$\mathrm{B}$''.
Because both $U_{f}(q)-U_{e}(q)$ and $U_{e}(q)-U_{g}(q)$ decrease in accordance with the wavepacket motion from $q=q_{\mathrm{i}}$ to $q=z^{\dagger }$, the ESA and SE peaks move toward the frequencies $\omega =U_{f}(q^{\dagger })-U_{e}(q^{\dagger })=E_{\mathrm{gap}}^{f-e}=5,000~\mathrm{cm}^{-1}$ and $U_{f}(q^{\dagger })-U_{e}(q^{\dagger })=E_{\mathrm{gap}}^{e-g}=1,000~\mathrm{cm}^{-1}$ as functions of time, respectively.
The intensity of the ESA signal decreases due to the de-excitation from $|\Phi _{e}(q)\rangle $ to $|\Phi _{g}(q)\rangle $.
The intensity of the SE signal also decreases, and becomes zero near $t=300~\mathrm{fs}$, because the SE from $|\Phi _{e}(q)\rangle $ and the absorption from $|\Phi _{g}(q)\rangle $ cancel each other.
Then, the absorption signal from the crossing region labeled ``$\mathrm{C}$'' appears and the position of the peak moves toward the resonant frequency of the product ground state labeled ``$\mathrm{D}$'', following the wavepacket motion.

In Fig.~\ref{fig:transient-absorption}(ii), the positive peak labeled ``$\mathrm{E}'$'' appears at $\omega =E_{\mathrm{gap}}^{f-e}$ after $t=400$ fs.
Its appearance is due to the violation of the positivity of the excited state population of $|\Phi _{e}(q^{\dagger })\rangle $; the $\mathrm{E}'$ peak in ESA signal arises from a negative population of $|\Phi _{e}\rangle $, which is observed as emission.
The negative population of $|\Phi _{e}(q^{\dagger })\rangle $ also appears as an absorptive contribution in the SE, and the intensity of the negative peak labeled ``$\mathrm{C}'$'' is enhanced.
When this negative population propagates to the $|\Phi _{g}(q^{\dagger })\rangle $ state, the absorption from this state is observed as the emission spectrum labeled ``$\mathrm{F}'$''.
These results indicate that the classical treatment does not have the capability to predict optical signals, in particular in the low-temperature regime.

In Figs.~\ref{fig:transient-absorption}(iii) and \ref{fig:transient-absorption}(iv), the calculated results from the phenomenological FFSH and Ehrenfest approaches are presented.
These results do not agree with the MS-LT-QSE results not only quantitatively but also qualitatively, in particular in the Ehrenfest case.
This is due to the poor estimation of non-adiabatic transition rate in these two approaches, as illustrated in Fig.~\ref{fig:population}.
Although the MS-SE approach exhibits unphysical emission or absorption peaks, the MS-SE result is closer to the accurate MS-LT-QSE results than the FSSH and Ehrenfest results.
This indicates that the MS-SE approach is physically more consistent than the FSSH and Ehrenfest approaches, even the positivity problem occurs.
Although the FSSH and Ehrenfest approaches are simple and easy to implement molecular dynamics simulations, the verification of the calculated results, in particular for calculation of nonlinear response function, should be made before comparing the experimental results.
Note that, when we calculate the effects of the non-adiabatic transition dynamic using the ensemble average of the electronic density matrix (i.e.~using the primary wavefunction given in Ref.~\onlinecite{tempelaar2013jcp}), the FSSH method produces a similar spectrum to the Ehrenfest result.

\section{CONCLUSION}
\label{sec:conclusion}
In this paper, we investigated (MS-)LT-QFPE and (MS-)LT-QSE that include QLT correction terms to satisfy the QFD theorem.
The (MS-)LT-QFPE and (MS-)LT-QSE were rigorously derived from path integral formalism.
We found that the (MS-)LT-QFPE can be used to obtain correct numerical descriptions of dynamics of a system coupled to an Ohmic bath when the QLT correction terms are treated properly, even in the strong coupling, low-temperature regime.
In the overdamped case, we can further reduce the momentum degrees of freedom from these equations, thereby obtaining the (MS-)LT-QSE.
Although the applicability of these equations is limited to the Ohmic case, they are significantly less computationally intensive than (MS-)QHFPE approaches in particular for the case of (MS-)LT-QSE.
Moreover, because structures of the PESs play essential roles in nonadiabatic transition phenomena, and because difference between the Markovian and non-Markovian noise cases is minor, the present formalism is sufficient for studying nonadiabatic transition phenomena.
Applications of this approach to the study of a molecular motor system will be presented in forthcoming papers.
The MS-LT-QFPE and MS-LT-QSE are also helpful for identifying purely quantum effects, because they allow us to compare the quantum results with the classical results obtained in the classical limit of the MS-LT-QFPE and MS-LT-QSE.

As shown in Appendix \ref{sec:appendx-qse}, although the LT-QSE accurately predicts the quantum dynamics in the case of strong friction, while we must truncate the number of QLT correction terms properly with estimating the timescales of each term given by, $\nu _{k}$, in comparison with the timescale of the system dynamics.
A correction of the LT-QSE using conventional QSE theories may suppress this ambiguity.
This is left for future investigations.

Because the (MS-)LT-QFPE is derived using a technique similar to that used in the conventional HEOM approach, an extension of the present formalism to the imaginary time formalism in calculations of the partition function should be straightforward \cite{tanimura2014jcp, tanimura2015jcp}.


\appendix
\section{Stochastic Liouville Description of (MS-)LT-QFPE and (MS-)LT-QSE}
\label{sec:appendx-continuous}
Because each contribution from the QLT terms is a Gaussian process, we can construct equations of motion in terms of continuous stochastic variables \cite{tanimura1989jpsj, tanimura2006jpsj}.
We introduce a set of stochastic variables $\vec{\Omega }\equiv (\dots ,\Omega _{k},\dots )$, where $\Omega _{k}$ is the auxiliary stochastic variable for the description of the $k$th QLT correction term.
Then, the Wigner distribution function is expressed as
\begin{subequations}
  \begin{align}
    \bm{W}^{\mathrm{d}}(p,q,\vec{\Omega },t)&\equiv \sum _{\vec{n}}\bm{W}_{\vec{n}}^{\mathrm{d}}(p,q,t)\phi _{\vec{n}}(\vec{\Omega }).
    \label{eq:lt-qfpe-brinkman-inv}
  \end{align}
  The inverse relation is expressed as
  \begin{align}
    \begin{split}
      \bm{W}^{\mathrm{d}}_{\vec{n}}(p,q,t)&=\int \!d\vec{\Omega }\,\bm{W}^{\mathrm{d}}(p,q,\vec{\Omega },t)\phi _{\vec{n}}^{(-1)}(\vec{\Omega }).
    \end{split}
    \label{eq:lt-qfpe-brinkman}
  \end{align}
\end{subequations}
Here, the functions $\phi _{\vec{n}}(\vec{\Omega })$ and $\phi _{\vec{n}}^{(-1)}(\vec{\Omega })$ are defined as
\begin{subequations}
  \begin{align}
    \phi _{\vec{n}}(\vec{\Omega })&\equiv \frac{\prod _{k}^{K}b_{k}^{n_{k}}}{\sqrt {\prod _{k}^{K}n_{k}!}}\psi _{\vec{0}}(\vec{\Omega })\psi _{\vec{n}}(\vec{\Omega })
    \label{eq:eigen-forward}
    \intertext{and}
    \phi _{\vec{n}}^{(-1)}(\vec{\Omega })&\equiv \frac{\sqrt {\prod _{k}^{K}n_{k}!}}{\prod _{k}^{K}b_{k}^{n_{k}}}\psi _{\vec{n}}(\vec{\Omega })\psi _{\vec{0}}(\vec{\Omega })^{-1},
    \label{eq:eigen-backward}
  \end{align}
\end{subequations}
where $\psi _{\vec{n}}(\vec{\Omega })$ is the Hermite function, $\psi _{\vec{n}}(\vec{\Omega })\equiv \prod _{k}^{K}\psi _{n_{k}}(\Omega _{k})$, with
\begin{align}
  \psi _{n_{k}}(\Omega _{k})&\equiv \frac{1}{\sqrt {\mathstrut 2^{n_{k}}n_{k}!\alpha _{k}\sqrt {\mathstrut \pi }}}H_{n_{k}}\left(\frac{\Omega _{k}}{\alpha _{k}}\right)\exp \left(-\frac{\Omega _{k}^{2}}{2\alpha _{k}^{2}}\right)
\end{align}
and the $n$th Hermite polynomial, $H_{n}(z)\equiv (-1)^{n}e^{z^{2}}(\partial ^{n}/\partial z^{n})e^{-z^{2}}$.
The coefficients $\alpha _{k}\neq 0$ and $b_{k}\neq 0$ are real numbers.
Then, the MS-WDF, $\bm{W}^{\mathrm{d}}(p,q,t)=\bm{W}_{\vec{0}}^{\mathrm{d}}(p,q,t)$, can be expressed as
\begin{align}
  \bm{W}^{\mathrm{d}}(p,q,t)=\int \!d\vec{\Omega }\,\bm{W}^{\mathrm{d}}(p,q,\vec{\Omega },t)
\end{align}
because of the orthogonality of the Hermite functions.

While the coefficients $\alpha _{k}$ and $b_{k}$ in Eqs.~\eqref{eq:eigen-forward} and \eqref{eq:eigen-backward} can be chosen in an arbitrary manner, we found that the choice $\alpha _{k}=\sqrt {\mathstrut 4\eta _{k}\zeta /\beta \hbar \omega _{0}\nu _{k}}$ and $b_{k}=\sqrt {\mathstrut 2}/\alpha _{k}\nu _{k}$ makes the equation of motion for $\bm{W}^{\mathrm{d}}(p,q,\vec{\Omega },t)$ simple.
Thus we obtain the stochastic Liouville description of the (MS-)LT-QFPE as
\begin{widetext}
  \begin{align}
    \begin{split}
      \frac{\partial }{\partial t}\bm{W}^{\mathrm{d}}(p,q,\vec{\Omega },t)
      &=-\biggl[\mathcal{L}_{\mathrm{qm}}(p,q)+\hat{\Xi }_{K}^{\mathrm{d}}(p,q)+\sum _{k}^{K}\left(\hat{\Phi }^{\mathrm{d}}(p,q)\hat{\Delta }_{k}(\Omega _{k})+\hat{\Xi }_{k}^{(\nu )}(\Omega _{k})\right)\biggr]\bm{W}^{\mathrm{d}}(p,q,\vec{\Omega },t)
      \label{eq:lt-qfpe-continuous},
    \end{split}
  \end{align}
\end{widetext}
where
\begin{subequations}
  \begin{align}
    \hat{\Delta }_{k}(\Omega _{k})&\equiv \nu _{k}\Omega _{k}+2\frac{\zeta }{\omega _{0}}\mathstrut \frac{2\eta _{k}}{\beta \hbar }\frac{\partial }{\partial \Omega _{k}}\\
    \intertext{and}
    \hat{\Xi }_{k}^{(\nu )}(\Omega _{k})&\equiv -\frac{\partial }{\partial \Omega _{k}}\left(\nu _{k}\Omega _{k}+\frac{\zeta }{\omega _{0}}\mathstrut \frac{2\eta _{k}}{\beta \hbar }\frac{\partial }{\partial \Omega _{k}}\right).
  \end{align}
\end{subequations}
For details of Eq.~\eqref{eq:lt-qfpe-continuous}, see Sec.~\nolink{\ref{sec:s-continuous-coordinate-representation}} of the Supporting Information.
Note that, while Eq.~\eqref{eq:lt-qfpe-continuous} is similar to the Fokker-Planck equation for classical non-Markovian dynamics via Markovian-type Fokker-Planck equations with ``virtual variables'' \cite{ferrario1979jmp, marchesoni1983jcp}, our variables $\vec{\Omega }=(\dots ,\Omega _{k},\dots )$ are introduced to describe the QLT correction terms from the Bose-Einstein distribution function.
Equation~\eqref{eq:lt-qfpe-brinkman} is similar to the discretized representation of a phase-space distribution of the classical Kramers equation (the Brinkman hierarchy) \cite{risken1989book, ikeda2017jcp}.
Therefore, the (MS-)LT-QFPE can be regarded as the Brinkman hierarchy representation of Eq.~\eqref{eq:lt-qfpe-continuous}.

Similarly, the (MS-)LT-QSE, Eq.~\eqref{eq:lt-qse-d}, is equivalent to the equation of motion,
\begin{widetext}
  \begin{align}
    \begin{split}
      \frac{\partial }{\partial t}\bm{f}^{\mathrm{d}}(q,\vec{\Omega },t)
      &=-\biggl[\mathcal{E}^{\mathrm{d}}(q)+\frac{\omega _{0}}{\zeta }\left(\mathcal{F}^{\mathrm{d}}(q)+\hat{\Xi }_{K}^{\mathrm{od},\mathrm{d}}(q)\right)
        +\sum _{k}^{K}\left(\hat{\Phi }^{\mathrm{od},\mathrm{d}}(q)\hat{\Delta }_{k}^{\mathrm{od}}(\Omega _{k})+\hat{\Xi }_{k}^{(\nu )\mathrm{od}}(\Omega _{k})\right)\biggr]\bm{f}^{\mathrm{d}}(q,\vec{\Omega },t)
      \label{eq:lt-qse-continuous},
    \end{split}
  \end{align}
\end{widetext}
where
\begin{subequations}
  \begin{align}
    \hat{\Delta }_{k}^{\mathrm{od}}(\Omega _{k})&\equiv \nu _{k}\Omega _{k}+2\frac{\omega _{0}}{\zeta }\frac{2\eta _{k}}{\beta \hbar }\frac{\partial }{\partial \Omega _{k}}
    \intertext{and}
    \hat{\Xi }_{k}^{(\nu )\mathrm{od}}(\Omega _{k})&\equiv -\frac{\partial }{\partial \Omega _{k}}\left(\nu _{k}\Omega _{k}+\frac{\omega _{0}}{\zeta }\frac{2\eta _{k}}{\beta \hbar }\frac{\partial }{\partial \Omega _{k}}\right).
  \end{align}
\end{subequations}
Further details of Eq.~\eqref{eq:lt-qse-continuous} are given in Sec.~\nolink{\ref{sec:s-continuous-coordinate-representation}} of the Supporting Information.

\section{Langevin Descriptions of LT-QFPE and LT-QSE}
\label{sec:appendx-langevin}
In the case that the system has only a single-state (i.e.~in the case of that $\bm{W}^{\mathrm{d}}(p,q,t)$ and $\bm{U}^{\mathrm{d}}(q)$ reduce to scaler functions, $W(p,q,t)$ and $U(q)$) and the quantum nature of the nuclear dynamics is weak (i.e.~anharmonicity of the potential is weak or friction is strong), higher-order terms of the Moyal product, Eq.~\eqref{eq:moyal-star}, can be omitted and Eq.~\eqref{eq:qm-potential-term} is approximated as
\begin{align}
  \mathcal{U}_{\mathrm{qm}}(p,q)W(p,q)\simeq \mathcal{U}_{\mathrm{cl}}(p,q)W(p,q)\equiv F(q)\frac{\partial }{\partial p}W(p,q).
  \label{eq:c-Liouvillian}
\end{align}
Here, we have introduced the force $F(q)\equiv -(1/\hbar )\partial U(q)/\partial q$.
In a harmonic potential case, the above expression is exact. 
Then, Eqs.~\eqref{eq:lt-qfpe-d} and \eqref{eq:lt-qfpe-continuous} can be decomposed into a set of Langevin equations,
\begin{subequations}
  \begin{align}
    \dot{q}(t)&=\omega _{0}p(t)
    \label{eq:langevin-q-raw},\\
    \dot{p}(t)&=F(q(t))-\zeta p(t)+\tilde{R}_{q}(t)+\sum _{k}^{K}\dot{\Omega }_{k}(t)
    \label{eq:langevin-p-raw},\\
    \intertext{and}
    \dot{\Omega }_{k}(t)&=-\nu _{k}\Omega _{k}(t)+\tilde{R}^{(\nu )}_{k}(t)
    \label{eq:langevin-w-k-raw}.
  \end{align}
\end{subequations}
Here, $\Omega _{k}(t)$ is an auxiliary stochastic variable for the $k$th index of $\vec{n}$, and $\tilde{R}_{q}(t)$ and $\tilde{R}^{(\nu )}_{k}(t)$ are Gaussian-white forces that satisfy the relations,
\begin{subequations}
  \begin{align}
    \langle \tilde{R}_{q}(t)\rangle &=\langle \tilde{R}_{k}^{(\nu )}(t)\rangle =0,\\
    \langle \tilde{R}_{q}(t)\tilde{R}_{q}(t')\rangle &=\frac{\zeta }{\omega _{0}}\frac{1}{\beta \hbar }\cdot 2\delta (t-t'),\\
    \langle \tilde{R}_{k}^{(\nu )}(t)\tilde{R}_{k}^{(\nu )}(t')\rangle &=\frac{\zeta }{\omega _{0}}\frac{2\eta _{k}}{\beta \hbar }\cdot 2\delta (t-t'),
  \end{align}
\end{subequations}
and $\langle \tilde{R}_{q}(t)\tilde{R}_{k}^{(\nu )}(t')\rangle =\langle \tilde{R}_{k}^{(\nu )}(t)\tilde{R}_{k}^{(\nu )}(t')\rangle =0$ ($k\neq k'$).
For details of Eqs.~\eqref{eq:langevin-q-raw}--\eqref{eq:langevin-w-k-raw}, see Sec.~\nolink{\ref{sec:s-langevin}} of the Supporting Information.
By introducing a quantum random force as
\begin{align}
  \tilde{R}_{\mathrm{qm}}(t)&=\tilde{R}_{q}(t)+\sum _{k}^{K}\dot{\Omega }_{k}(t),
\end{align}
the set of Langevin equations, Eqs.~\eqref{eq:langevin-q-raw}, \eqref{eq:langevin-p-raw}, and \eqref{eq:langevin-w-k-raw}, can be rewritten as
\begin{subequations}
  \begin{align}
    \dot{q}(t)&=\omega _{0}p(t)
    \label{eq:langevin-q}\\
    \intertext{and}
    \dot{p}(t)&=F(q(t))-\zeta p(t)+\tilde{R}_{\mathrm{qm}}(t)
    \label{eq:langevin-p}.
  \end{align}
\end{subequations}
The random force $\tilde{R}_{\mathrm{qm}}(t)$ satisfies the QFD theorem as
\begin{subequations}
  \begin{align}
    &\langle \tilde{R}_{\mathrm{qm}}(t)\tilde{R}_{\mathrm{qm}}(t')\rangle =C_{K}(t-t')\\
    &\mathop{\rightarrow }_{K\rightarrow \infty }2\int \!d\omega \,\frac{\zeta }{\omega _{0}}\omega \left(n(\omega )+\frac{1}{2}\right)\cos \omega (t-t').
  \end{align}
\end{subequations}
The set of equations~\eqref{eq:langevin-q} and \eqref{eq:langevin-p} are the quantum Langevin equation \cite{ford1988pra, weiss2011book} for the c-number variables, $p(t)$ and $q(t)$ (i.e.~the quasi-classical Langevin equation \cite{senitzky1960pr, schmid1982jltp}).
In the cases that anharmonicity of the system is strong, we cannot employ the above equations.
In such cases, we have to evaluate the quantum Langevin equation described by the operators, $\hat{p}(t)$ and $\hat{q}(t)$, or to employ Wigner description (i.e.~Eqs.~\eqref{eq:lt-qfpe-d} and \eqref{eq:lt-qfpe-continuous}) with the quantum Liouvillian \eqref{eq:qm-Liouvillian}.
This indicates that, the LT-QFPE, Eqs.~\eqref{eq:lt-qfpe-d} and \eqref{eq:lt-qfpe-continuous}, can be regarded as the Fokker-Planck equations equivalent to the quantum Langevin equation.
Note that, while Eqs.~\eqref{eq:langevin-q-raw}, \eqref{eq:langevin-p-raw}, and \eqref{eq:langevin-w-k-raw} are similar to the generalized Langevin equation for classical non-Markovian dynamics via Markovian-type Lanvegin equations with virtual variables \cite{ferrario1979jmp, marchesoni1983jcp, martens2002jcp, kawai2015jcp}, our stochastic variables $\vec{\Omega }=(\dots ,\Omega _{k},\dots )$ are introduced to describe the QLT correction terms from the Bose-Einstein distribution function.

It should be noted that, for the LT-QFPE with a single PES, Eq.~\eqref{eq:lt-qse-d} can also be decomposed into a set of Langevin equations,
\begin{subequations}
  \begin{align}
    \frac{\zeta }{\omega _{0}}\dot{q}(t)&=F(q(t))+\tilde{R}_{q}(t)+\sum _{k}^{K}\dot{\Omega }_{k}(t)
    \label{eq:ov-langevin-q-raw}\\
    \intertext{and}
    \dot{\Omega }_{k}(t)&=-\nu _{k}\Omega _{k}+\tilde{R}^{(\nu )}_{k}(t)
    \label{eq:ov-langevin-w-k-raw}.
  \end{align}
\end{subequations}
By introducing $\tilde{R}_{\mathrm{qm}}(t)$, the set of Langevin equations, Eqs.~\eqref{eq:ov-langevin-q-raw} and \eqref{eq:ov-langevin-w-k-raw}, can be rewritten as
\begin{align}
  \frac{\zeta }{\omega _{0}}\dot{q}(t)&=F(q(t))+\tilde{R}_{\mathrm{qm}}(t).
\end{align}
This is the ``overdamped'' quantum Langevin equation for the c-number variable, $q(t)$ (i.e.~the inertia term $\propto \ddot{q}(t)$ is omitted).
For details of Eqs.~\eqref{eq:ov-langevin-q-raw} and \eqref{eq:ov-langevin-w-k-raw}, see Sec.~\nolink{\ref{sec:s-langevin}} of the Supporting Information.

\section{Comparison of QSE and LT-QSE}
\label{sec:appendx-qse}
In this appendix, we compare our LT-QSE theory with conventional QSE theories.

The QSE is proposed in Refs.~\onlinecite{philip2000adp, ankerhold2001prl, ankerhold2008prl}.
It is given by
\begin{align}
  \frac{\partial }{\partial t}f(q,t)&=\frac{\omega _{0}}{\zeta }\frac{\partial }{\partial q}\left[\frac{1}{\hbar }\frac{\partial U(q)}{\partial q}+\frac{1}{\beta \hbar }\left(1+\beta \lambda \frac{\partial ^{2}U(q)}{\partial q^{2}}\right)\frac{\partial }{\partial q}\right]f(q,t)
  \label{eq:qse-gen}
\end{align}
in terms of our dimensionless coordinate $q$.
Here, we have
\begin{align}
  \lambda &\equiv \frac{\omega _{0}}{\pi \zeta }\log \left(\frac{\beta \hbar \zeta }{2\pi }\right).
\end{align}
In the case of the harmonic oscillator~\eqref{eq:harmonic-PES}, Eq.~\eqref{eq:qse-gen} can be written
\begin{align}
  \frac{\partial }{\partial t}f(q,t)&=\frac{\omega _{0}^{2}}{\zeta }\frac{\partial }{\partial q}\left[q+\frac{1}{\beta \hbar \omega _{0}}\left(1+\beta \hbar \omega _{0}\lambda \right)\frac{\partial }{\partial q}\right]f(q,t).
  \label{eq:qse}
\end{align}
Better quantum corrections for the QSE are presented in Refs.~\onlinecite{maier2010pre, maier2010pre2}.
For a harmonic potential, this becomes
\begin{align}
  \frac{\partial }{\partial t}f(q,t)&=\Omega \frac{\partial }{\partial q}\left(q+\langle q^{2}\rangle _{\beta ,\zeta }\frac{\partial }{\partial q}\right)f(q,t),
  \label{eq:qse_prime}
\end{align}
where $\Omega \equiv \zeta /2-\sqrt {\mathstrut (\zeta /2)^{2}-\omega _{0}^{2}}$, and $\langle q^{2}\rangle _{\beta ,\zeta }$ is given in Eq.~\eqref{eq:analytical-q-variance}.
Hereafter, we refer to Eqs.~\eqref{eq:qse} and \eqref{eq:qse_prime} as the QSE and QSE', respectively.

\begin{figure}
  \centering
  \includegraphics[scale=\SingleColFigScale]{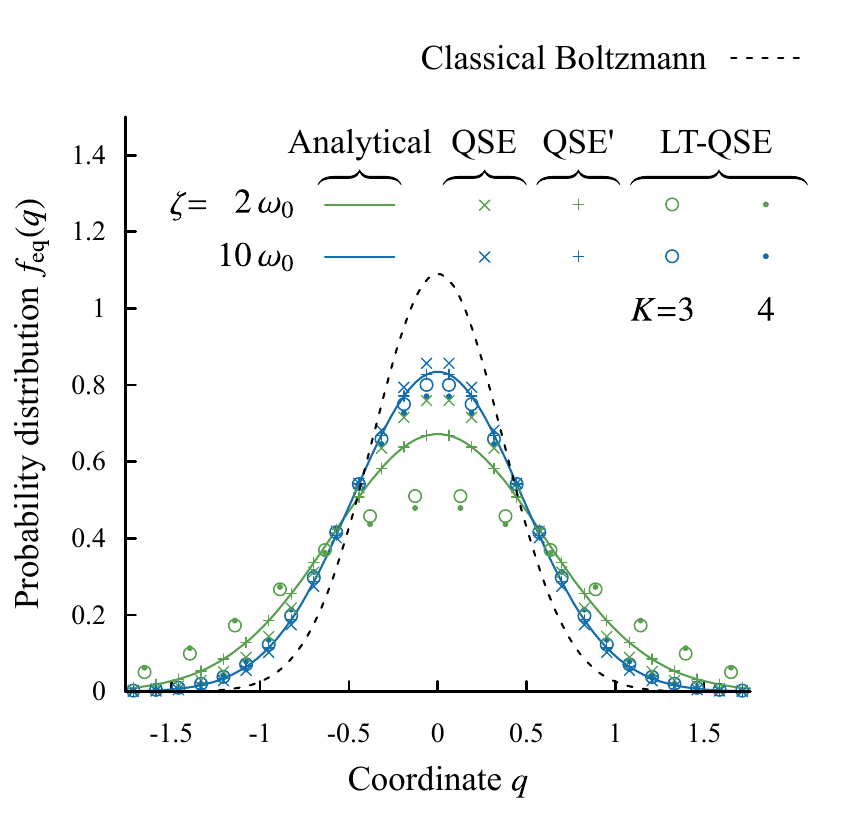}
  \caption{
    The equilibrium distribution, $f_{\mathrm{eq}}(q)$, calculated with the QSE, QSE' and LT-QSE for the critically-damped, $\zeta =2\,\omega _{0}$ (green), and overdamped, $\zeta =10\, \omega _{0}$ (blue), cases at low temperature, $\beta \hbar \omega _{0}=7.47$.
    The classical (dashed) and quantum (solid) equilibrium distributions were obtained from the analytical expression in Eq.~\eqref{eq:analytical-distribution}.
  }
  \label{fig:qdistribution-SE}
\end{figure}
In Fig.~\ref{fig:qdistribution-SE}, we display the steady-state solutions of the QSE, QSE', and LT-QSE for several values of the damping strength at low temperature, $\beta \hbar \omega _{0}=7.47$.
The other parameter values are the same as in Sec.~\ref{sec:brownian}.
In the overdamped case, all calculated results are qualitatively similar to the analytical result.
This is because the QSE and QSE' are constructed so as to reproduce the steady-state solution.
The QSE', in particular, utilizes $\langle q^{2}\rangle _{\beta ,\zeta }$, for this reason it reproduces the analytical result even in the critical-damping case.

Note that if we assume the overdamped approximation Eq.~\eqref{eq:analytical-solution-overdamped} for the analytical equilibrium distribution Eq.~\eqref{eq:analytical-distribution} instead of Eq.~\eqref{eq:analytical-solution}, the integral in Eq.~\eqref{eq:analytical-q-variance} diverges under the infinite summation of the Matsubara frequencies as
\begin{align}
  \langle q^{2}\rangle _{\beta ,\zeta }^{\zeta \gg \omega _{0}}&=\frac{\tilde{\gamma }}{\beta \hbar \omega _{0}}\sum _{k=-\infty }^{\infty }\frac{1}{\tilde{\gamma }+\left|\tilde{\nu }_{k}\right|}\rightarrow \infty .
\end{align}
The calculated results using the LT-QSE theory become close to the above analytical result by employing larger hierarchical space (i.e.~by increasing $K$). Thus, the steady-state solution predicted by the LT-QSE theory deviates as $K$ increases.
This divergence is similar to the ultra-violet divergence of $\langle p^{2}\rangle $ .
This indicates that, in order to calculate physical quantities on the basis of LT-QSE, we must truncate the number of QLT correction terms properly by estimating the timescales of them, $\nu _{k}$, in comparison with the timescale of the system dynamics.
Note that a Drude spectral density model for reduced electronic states has also this problem, because Eq.~\eqref{eq:analytical-solution-overdamped} is equivalent to the symmetrized correlation function of the collective noise coordinate in the Drude spectral density model.

\begin{figure}
  \centering
  \includegraphics[scale=\SingleColFigScale]{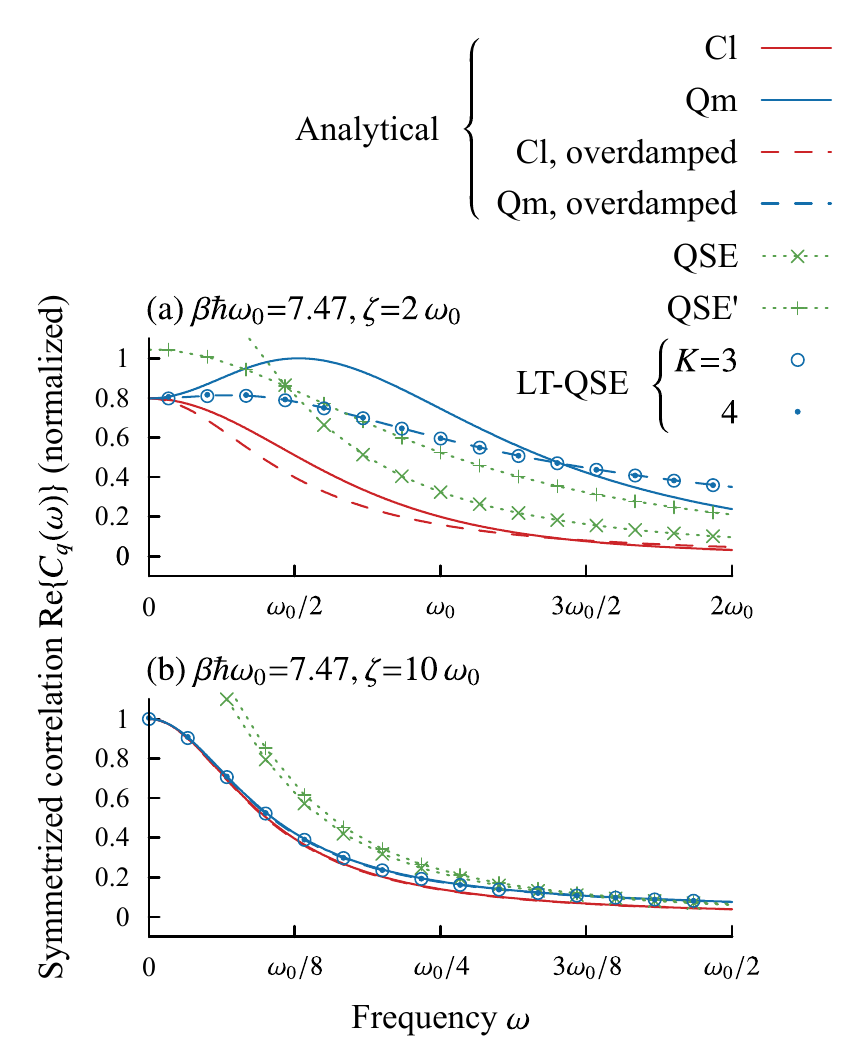}
  \caption{
    The symmetrized correlation functions, $C(\omega )$, calculated with the QSE, QSE', and LT-QSE for the (a) critically-damped and (b) overdamped cases at low temperature, $\beta \hbar \omega _{0}=7.47$.
    The solid and dashed curves represent the analytically derived solution, Eq.~\eqref{eq:analytical-solution}, and its overdamped limit, Eq.~\eqref{eq:analytical-solution-overdamped}, respectively.
    The red and blue curves were obtained with classical and quantum treatments of $\coth (\beta \hbar \omega /2)$.
    The values are normalized with respect to the maximum of Eqs.~\eqref{eq:analytical-solution}.
  }
  \label{fig:1D-SE}
\end{figure}
In Fig.~\ref{fig:1D-SE}, we display the symmetrized correlation functions calculated with the QSE, QSE' and LT-QSE under the same conditions as in the case of Fig.~\ref{fig:qdistribution-SE}.
Because the QSE and QSE' attempt to account for all quantum corrections with Markovian terms ignoring the non-Markovian nature of the quantum noise, they cannot reproduce the dynamics at low temperature.
Contrastingly, the calculated results with the LT-QSE theory become close to the exact solution in the overdamped case, Eq.~\eqref{eq:analytical-solution}, by increasing $K$, and both numerical results and Eq.~\eqref{eq:analytical-solution} approach the solution Eq.~\eqref{eq:analytical-solution-overdamped} for large $\zeta $.

\section{Zusman Equation}
\label{sec:appendx-Zusman}
In this appendix, we show the relation between the present equation and Zusman equation.
For simplicity, we assume a two-level system described by the diabatic harmonic PESs as
\begin{subequations}
  \begin{align}
    \bm{U}^{\mathrm{d}}(q)&=\bm{A}-\hbar d\omega _{0}\bm{B}q+\frac{\hbar \omega _{0}}{2}q^{2}\\
    \intertext{and}
    \bm{F}^{\mathrm{d}}(q)&=-\frac{1}{\hbar }\frac{\partial \bm{U}^{\mathrm{d}}(q)}{\partial q}=d\omega _{0}\bm{B}-\omega _{0}q,
  \end{align}
\end{subequations}
with matrices
\begin{align}
  \bm{A}&=
  \begin{pmatrix}
    0 & V\\
    V & E_{0}
  \end{pmatrix}
  &\text{and}&&
  \bm{B}&=
  \begin{pmatrix}
    0 & 0\\
    0 & 1
  \end{pmatrix}.
\end{align}
Here, $V$ is the diabatic coupling constant between $|0\rangle $ and $|1\rangle $, $d$ is the dimensionless displacement between the minima of two PESs, and $E_{0}\equiv \Delta E+\lambda $ is the summation of the driving force $\Delta E$ and the reorganization energy $\lambda \equiv \hbar \omega _{0}d^{2}/2$.
In this case, the MS-SE can be rewritten as
\begin{align}
  \begin{split}
    \frac{\partial }{\partial t}\bm{f}^{\mathrm{d}}(q,t)
    &=-\frac{i}{\hbar }\bm{A}^{\times }\bm{f}^{\mathrm{d}}(q,t)+d\omega _{0}qi\bm{B}^{\times }\bm{f}^{\mathrm{d}}(q,t)\\
    &\quad -\tilde{\gamma }d\frac{\partial }{\partial q}\frac{1}{2}\bm{B}^{\circ }\bm{f}^{\mathrm{d}}(q,t)+\hat{\Gamma }(q)\bm{f}^{\mathrm{d}}(q,t).
  \end{split}
  \label{eq:msse-harmonic}
\end{align}
Here, we have introduced commutation and anti-commutation hyper operators, $\bm{O}^{\times /\circ }\bm{f}\equiv \bm{O}\bm{f}\mp \bm{f}\bm{O}$, and
\begin{align}
  \hat{\Gamma }(q)\equiv \tilde{\gamma }\frac{\partial }{\partial q}\left(q+\frac{1}{\beta \hbar \omega _{0}}\frac{\partial }{\partial q}\right).
\end{align}
Note that Eq.~\eqref{eq:msse-harmonic} is equivalent to the stochastic Liouville representation of the HEOM for a Drude spectral density in the high-temperature limit, given in Refs.~\onlinecite{tanimura1989jpsj, shi2009jcp2}.
When we neglect the electronic-nuclear interaction from the force term (i.e.~$\bm{F}^{\mathrm{d}}(q)\rightarrow -\omega _{0}q$ and the anti-commutation term ($\propto \bm{B}^{\circ }$) is omitted from Eq.~\eqref{eq:msse-harmonic}), we obtain the ZE in a matrix representation as
\begin{align}
  \begin{split}
    \frac{\partial }{\partial t}\bm{f}^{\mathrm{d}}(q,t)
    &=-\frac{i}{\hbar }\bm{A}^{\times }\bm{f}^{\mathrm{d}}(q,t)+d\omega _{0}qi\bm{B}^{\times }\bm{f}^{\mathrm{d}}(q,t)+\hat{\Gamma }(q)\bm{f}^{\mathrm{d}}(q,t).
  \end{split}
  \label{eq:zusman-matrix-rep}
\end{align}
The equations of motion for diabatic matrix elements are expressed as
\begin{widetext}
  \begin{subequations}
    \begin{align}
      &\begin{aligned}
         \frac{\partial }{\partial t}f_{00}^{\mathrm{d}}(q,t)&=-\frac{i}{\hbar }(f_{10}^{\mathrm{d}}(q,t)-f_{01}^{\mathrm{d}}(q,t))V+\hat{\Gamma }(q)f_{00}^{\mathrm{d}}(q,t),
       \end{aligned}\\
      &\begin{aligned}
         \frac{\partial }{\partial t}f_{11}^{\mathrm{d}}(q,t)&=-\frac{i}{\hbar }(f_{01}^{\mathrm{d}}(q,t)-f_{10}^{\mathrm{d}}(q,t))V+\hat{\Gamma }(q)f_{11}^{\mathrm{d}}(q,t),
       \end{aligned}
      \intertext{and}
      &\begin{aligned}
         \frac{\partial }{\partial t}f_{01}^{\mathrm{d}}(q,t)&=-\frac{i}{\hbar }f_{01}^{\mathrm{d}}(q,t)(\hbar \omega _{0}dq-E_{0})
         -\frac{i}{\hbar }(f_{11}^{\mathrm{d}}(q,t)-f_{00}^{\mathrm{d}}(q,t))V+\hat{\Gamma }(q)f_{01}^{\mathrm{d}}(q,t).
       \end{aligned}
    \end{align}
  \end{subequations}
\end{widetext}
The above equations are equivalent to the original ZE in Ref.~\onlinecite{zusman1980cp} except shift and scaling factors of $q$.
As this derivation indicates, the MS-SE can be regarded as a generalization of the ZE for arbitrary PESs with including the quantum dynamical effects arising from the electric states.
Note that, the ZE for harmonic PESs including this dynamical effect is given in Refs.~\onlinecite{garg1985jcp, shi2009jcp2}.
This is also sometimes referred to as the ZE.

\section{Adiabatic Representations of MS-LT-QFPE and MS-LT-QSE}
\label{eq:app-adiabatic-rep}
In this appendix, we present the adiabatic representations of MS-LT-QFPE \eqref{eq:lt-qfpe-d} and MS-LT-QSE \eqref{eq:lt-qse-d}.
  
We can rewrite Eqs.~\eqref{eq:adiabatic-rep} and \eqref{eq:adiabtic-transform} in terms of the Moyal star product, Eq.~\eqref{eq:moyal-star}, as
\begin{align}
  \bm{W}^{\mathrm{a}}(p,q,t)&=\bm{Z}\left(q\right)^{\dagger }{\star }\bm{W}^{\mathrm{d}}(p,q,t){\star }\bm{Z}\left(q\right).
  \label{eq:W-adiabatic-transform}
\end{align}
Inserting this into Eq.~\eqref{eq:lt-qfpe-d}, we obtain the MS-QFP-LT in the adiabatic representation as
\begin{widetext}
  \begin{align}
    \begin{split}
      \frac{\partial }{\partial t}\bm{W}_{\vec{n}}^{\mathrm{a}}(p,q,t)
      &=-\Bigl(\mathcal{L}_{\mathrm{qm}}^{\mathrm{a}}(p,q)+\sum _{k}^{K}n_{k}\gamma _{k}+\hat{\Xi }_{\mathrm{qm}}^{\mathrm{a}}(p,q)\Bigr)\bm{W}_{\vec{n}}^{\mathrm{a}}(p,q,t)\\
      &\quad -\sum _{k}^{K}\hat{\Phi }^{\mathrm{a}}(p,q)\bm{W}_{\vec{n}+\vec{e}_{k}}^{\mathrm{a}}(p,q,t)
      -\sum _{k}^{K}n_{k}\gamma _{k}\hat{\Theta }_{k}^{\mathrm{a}}(p,q)\bm{W}_{\vec{n}-\vec{e}_{k}}^{\mathrm{a}}(p,q,t),
      \label{eq:mshqfpe-a}
    \end{split}
  \end{align}
\end{widetext}
\begin{subequations}
  where
  \begin{align}
    \mathcal{L}_{\mathrm{qm}}^{\mathrm{a}}(p,q)&\equiv \mathcal{K}(p,q)+\mathcal{U}_{\mathrm{qm}}^{\mathrm{a}}(p,q)+\mathcal{D}_{\mathrm{qm}}^{\mathrm{a}}(p,q)
  \end{align}
  is the quantum Liouvillian for the MSWDF in the adiabatic representation, and we have
  \begin{align}
    \begin{split}
      \mathcal{U}_{\mathrm{qm}}^{\mathrm{a}}(p,q)\bm{W}^{\mathrm{a}}(p,q)&\equiv \frac{i}{\hbar }\bm{U}^{\mathrm{a}}(q)^{\times _{(\star )}}\bm{W}^{\mathrm{a}}(p,q)
    \end{split}
  \end{align}
  and
  \begin{widetext}
    \begin{align}
      \begin{split}
        \mathcal{D}_{\mathrm{qm}}(p,q)\bm{W}^{\mathrm{a}}(p,q)
        &\equiv -i\frac{\omega _{0}}{2}\bm{d}(q)^{\circ _{(\star )}}\frac{\partial \bm{W}^{\mathrm{a}}(p,q)}{\partial q}+\omega _{0}\bm{d}(q)^{\times _{(\star )}}p\bm{W}^{\mathrm{a}}(p,q)
        -\frac{i\omega _{0}}{2}\left(\bm{h}(q){\star }\bm{W}^{\mathrm{a}}(p,q)-\bm{W}^{\mathrm{a}}(p,q){\star }\bm{h}(q)^{\dagger }\right).
      \end{split}
    \end{align}
  \end{widetext}
\end{subequations}
Here, we have introduced the commutation and anti-commutation hyper operators with the Moyal product, $\bm{O}^{\times /\circ _{(\star )}}\bm{f}\equiv \bm{O}\star \bm{f}\mp \bm{f}\star \bm{O}$.
The non-Markovian noise terms are $\hat{\Phi }^{\mathrm{a}}(p,q)\equiv \hat{\Phi }^{\mathrm{d}}(p,q)$ and $\hat{\Theta }_{k}^{\mathrm{a}}(p,q)\equiv \hat{\Theta }_{k}^{\mathrm{d}}(p,q)$, and the Markovian noise term, Eq.~\eqref{eq:xi-d}, becomes
\begin{widetext}
  \begin{align}
    \begin{split}
      \hat{\Xi }_{\mathrm{qm}}^{\mathrm{a}}(p,q)\bm{W}^{\mathrm{a}}(p,q)
      &\equiv -\frac{\zeta }{\omega _{0}}\frac{\partial }{\partial p}\biggl(\omega _{0}p\bm{W}^{\mathrm{a}}(p,q)-\frac{i\omega _{0}}{2}\bm{d}\left(p,q\right)^{\circ _{(\star )}}\bm{W}^{\mathrm{a}}(p,q)
      +\frac{1}{\beta \hbar }\frac{\partial }{\partial p}\bm{W}^{\mathrm{a}}(p,q)\biggr)
      +\sum _{k}^{K}\hat{\Phi }^{\mathrm{a}}(p,q)\hat{\Theta }_{k}^{\mathrm{a}}(p,q)\bm{W}^{\mathrm{a}}(p,q).
    \end{split}
  \end{align}
\end{widetext}

In the Smoluchowski limit, we introduce the adiabatic representation of $\bm{f}^{\mathrm{d}}(q)$:
\begin{align}
  \bm{f}^{\mathrm{a}}(q,t)=\bm{Z}\left(q\right)^{\dagger }\bm{f}^{\mathrm{d}}(q,t)\bm{Z}\left(q\right).
  \label{eq:f-adiabatic-transform}
\end{align}
Here, we have omitted the higher-order contributions from the Moyal product in Eq.~\eqref{eq:W-adiabatic-transform}, because such contributions from the quantum coherence, $z-z'$, have been removed in the Smoluchowski limit.
Then the MS-LT-QSE \eqref{eq:lt-qse-d} becomes
\begin{widetext}
  \begin{align}
    \begin{split}
      \frac{\partial }{\partial t}\bm{f}_{\vec{n}}^{\mathrm{a}}(q,t)
      &=-\left[\mathcal{E}^{\mathrm{a}}(q)+\sum _{k}^{K}n_{k}\gamma _{k}+\frac{\omega _{0}}{\zeta }\Bigl(\mathcal{F}^{\mathrm{a}}(q)+\hat{\Xi }_{K}^{\mathrm{od},\mathrm{a}}(q)\Bigr)\right]\bm{f}_{\vec{n}}^{\mathrm{a}}(q,t)\\
      &\quad -\sum _{k}^{K}\hat{\Phi }^{\mathrm{od},\mathrm{a}}(q)\bm{f}_{\vec{n}+\vec{e}_{k}}^{\mathrm{a}}(q,t)
      -\frac{\omega _{0}}{\zeta }\sum _{k}^{K}n_{k}\gamma _{k}\hat{\Theta }_{k}^{\mathrm{od},\mathrm{a}}(q)\bm{f}_{\vec{n}-\vec{e}_{k}}^{\mathrm{a}}(q,t),
    \end{split}
    \label{eq:lt-qse-a}
  \end{align}
  where the operators appearing in Eq.~\eqref{eq:lt-qse-d} are transformed as
  \begin{subequations}
    \begin{align}
      \mathcal{E}^{\mathrm{a}}(q)\bm{f}(q)&\equiv \frac{i}{\hbar }\bm{U}^{\mathrm{a}}(q)^{\times }\bm{f}(q),\\
      \mathcal{F}^{\mathrm{a}}(q)\bm{f}^{\mathrm{a}}(q)&\equiv \frac{1}{2}\bm{F}^{\mathrm{a}}(q)^{\circ }\left(\frac{\partial }{\partial q}+\bm{d}(q)^{\times }\right)\bm{f}^{\mathrm{a}}(q)+\frac{1}{2}\bm{A}^{\mathrm{a}}(q)^{\circ }\bm{f}^{\mathrm{a}}(q),
    \end{align}
  \end{subequations}
  \begin{subequations}
    \begin{align}
      \hat{\Phi }^{\mathrm{od},\mathrm{a}}(q)\bm{f}^{\mathrm{a}}(q)&\equiv -\left(\frac{\partial }{\partial q}+\bm{d}(q)^{\times }\right)\bm{f}^{\mathrm{a}}(q),\\
      \hat{\Theta }_{k}^{\mathrm{od},\mathrm{a}}(q)\bm{f}^{\mathrm{a}}(q)&\equiv \frac{2\eta _{k}}{\beta \hbar }\left(\frac{\partial }{\partial q}+\bm{d}(q)^{\times }\right)\bm{f}^{\mathrm{a}}(q),\\
      \intertext{and}
      \hat{\Xi }_{K}^{\mathrm{od},\mathrm{a}}(q)\bm{f}^{\mathrm{a}}(q)&\equiv -\frac{1}{\beta \hbar }\left(\frac{\partial }{\partial q}+\bm{d}(q)^{\times }\right)^{2}\bm{f}^{\mathrm{a}}(q)+\sum _{k}^{K}\hat{\Phi }^{\mathrm{od},\mathrm{a}}(q)\hat{\Theta }_{k}^{\mathrm{od},\mathrm{a}}(q).
    \end{align}
  \end{subequations}
\end{widetext}
Here, we have introduced the force acting on the adiabatic states,
\begin{align}
  \begin{split}
    \bm{F}^{\mathrm{a}}(q)&\equiv \bm{Z}(q)^{\dagger }\bm{F}^{\mathrm{d}}(q)\bm{Z}(q)\\
    &=-\frac{1}{\hbar }\frac{\partial \bm{U}^{\mathrm{a}}(q)}{\partial q}-\frac{1}{\hbar }\bigl(\bm{d}(q)\bm{U}^{\mathrm{a}}(q)-\bm{U}^{\mathrm{a}}(q)\bm{d}(q)\bigr)
  \end{split}
\end{align}
and its derivative,
\begin{align}
  \begin{split}
    \bm{A}^{\mathrm{a}}(q)&\equiv \bm{Z}(q)^{\dagger }\frac{\partial \bm{F}^{\mathrm{d}}(q)}{\partial q}\bm{Z}(q)
    =\frac{\partial \bm{F}^{\mathrm{a}}(q)}{\partial q}+\bm{d}(q)\bm{F}^{\mathrm{a}}(q)-\bm{F}^{\mathrm{a}}\bm{d}(q).
  \end{split}
\end{align}

\begin{acknowledgement}
  This work was supported by JSPS KAKENHI Grant Number JP26248005.
\end{acknowledgement}

\begin{suppinfo}
  \begin{description}
  \item[ct8b01195\_si\_001.zip]
    The C++ source codes that we developed, which allow for the treatment of the phase and coordinate space dynamics with any single-state or multi-state potential forms.
  \item[ct8b01195\_si\_002.pdf]
    The derivations of the (MS-)LT-QFPE and (MS-)LT-QSE, the derivation of the stochastic Liouville and Langevin descriptions of (MS-)LT-QFPE and (MS-)LT-QSE, the truncation scheme of hierarchy, note on the ultra-violent divergence, and the details of the numerical calculations.
  \end{description}
  The Supporting Information is available free of charge on the ACS Publications website at DOI: \href{https://doi.org/10.1021/acs.jctc.8b01195}{10.1021/acs.jctc.8b01195}.
\end{suppinfo}
  
\let\emph=\textit
\bibliography{ikeda_JCTC2018rev.bib}

\end{document}